\documentclass[12pt]{iopart}
\expandafter\let\csname equation*\endcsname\relax
\expandafter\let\csname endequation*\endcsname\relax
\usepackage{amsmath,amsfonts,amssymb}
\usepackage{bm}
\usepackage{bbm}
\usepackage{cite}
\usepackage{graphicx} 
\usepackage[colorlinks, linkcolor=blue, citecolor = blue, urlcolor = blue]{hyperref}
\usepackage{booktabs}

\newcommand \kv {\mathbf{k}}
\newcommand \corr {C_2}

\begin{document}

\title[Gradient elasticity in Swift-Hohenberg and phase-field crystal models]
{Gradient elasticity in Swift-Hohenberg and phase-field crystal models}

\author{Lucas Benoit-{}-Mar\'echal$^1$, and Marco Salvalaglio$^{1,2}$}
\address{$^1$Institute of Scientific Computing, TU Dresden, 01062 Dresden, Germany.}
\address{$^2$Dresden Center for Computational Materials Science, TU Dresden, 01062 Dresden, Germany.}
\ead{marco.salvalaglio@tu-dresden.de}
\vspace{10pt}

\begin{abstract}
The Swift-Hohenberg (SH) and Phase-Field Crystal (PFC) models are minimal yet powerful approaches for studying phenomena such as pattern formation, collective order, and defects via smooth order parameters. They are based on a free-energy functional that inherently includes elasticity effects. This study addresses how gradient elasticity (GE), a theory that accounts for elasticity effects at microscopic scales by introducing additional characteristic lengths, is incorporated into SH and PFC models. After presenting the fundamentals of these theories and models, we first calculate the characteristic lengths for various lattice symmetries in an approximated setting. We then discuss numerical simulations of stress fields at dislocations and comparisons with analytic solutions within first and second strain-gradient elasticity. Effective GE characteristic lengths for the elastic fields induced by dislocations are found to depend on the free-energy parameters in the same manner as the phase correlation length, thus unveiling how they change with the quenching depth. The findings presented in this study enable a thorough discussion and analysis of small-scale elasticity effects in pattern formation and crystalline systems using SH and PFC models and, importantly, complete the elasticity analysis therein. Additionally, we provide a microscopic foundation for GE in the context of order-disorder phase transitions.
\end{abstract}


\newpage
\section{Introduction}
\label{sec:intro}

Gradient elasticity (GE) is a theoretical framework that extends classical continuum mechanics with additional higher-order derivatives to capture the effects of small length scales \cite{toupin1962elastic,toupin1964theories,Mindlin1964,Mindlin1968,ASKES20111962,Lazar2018jmmp,Lazar2022_iso}. It incorporates additional characteristic lengths accounting for internal material structure and discreteness, thereby allowing the description of size effects relevant at the nano- and micro-scale. Such a theory has been tested and exploited in different frameworks, e.g., classical mechanics and dislocation dynamics, and applied in contexts such as flexoelectricity and biomechanics \cite{LamJMPS2003,ZHANG20053833,GAO20077486,akgoz2011application,PO2014161,mao2014insights,Jiang2022}. It has also been extended to describe strain-gradient plasticity \cite{fleck1994strain,gao1999mechanism,hutchinson1997strain,fleck2001reformulation}.
Parameters entering the theory have been connected to interatomic potentials and, in general, can be determined by atomistic methods \cite{shodja2013ab,ADMAL201793,po2019green}. A key feature of GE is that it overcomes some limitations of continuum elasticity theory, such as nonphysical singularities emerging from approximated descriptions of small scales. This leads, for instance, to regularized elastic fields at the dislocation core \cite{Lazar2005,LAZAR20061787,Lazar2013,Lazar2022,lazar2024}, which are instead singular in classic linear elasticity \cite{anderson2017}.

Elasticity-driven pattern formation emerges in a variety of physical systems---e.g., hard and soft materials, biological tissues \cite{MULLER2004157,B516741H}---due to the interplay between mechanical deformation and other physical contributions such as capillarity, solidification, crystal growth dynamics, and grand potential jumps across interfaces.
In the study of general features of pattern formation, a central role has been played by the so-called Swift-Hohenberg (SH) model \cite{Swift1977,Cross1993} and its extensions. It describes the spatiotemporal evolution of a real order parameter field $\psi$ based on a partial differential equation comprising linear differential operators and polynomial terms (the SH equation). Such an equation favors the linear growth of periodic modes of $\psi$ until the nonlinear contributions induce saturation, leading to the selection of patterns with different symmetries depending on the parametrization. Notably, the SH equation can be written in a variational form through the definition of a free energy (or Lyapunov) functional $F[\psi]$ as $\partial_t \psi=-\delta F / \delta \psi$. Therefore, stable stationary states correspond to the minima of such functional. 
While it was initially proposed to model the so-called Rayleigh-B\'enard convection in a heated fluid, similar equations have been exploited to describe the emergence of patterns involving elasticity effects in different contexts \cite{aranson2006patterns,stoop2015curvature,oza2016generalized}.
The connection of SH-like models to elasticity became evident in the conservative formulation, the so-called phase-field crystal (PFC) model, explicitly introduced to model elasticity in crystal growth \cite{Elder2002,Elder2004,Provatas2010,Emmerich2012}. 
In its basic formulation, the PFC model is based on a free energy analogous to the SH model, albeit with conservative dynamics $\partial_t \psi=\nabla^2 (\delta F / \delta \psi)$, where $\psi$ can be considered as the time average of microscopic atomic density over vibrational timescales. The periodic pattern, in that case, represents the lattice structure of crystalline materials. Such a model has been successfully applied to describe mesoscale phenomena in crystals, including elasticity effects during crystal growth, dislocation dynamics, and microstructure evolution \cite{Elder2004,Emmerich2012,Berry2014,Backofen14,GRANASY2019}.

Elasticity is thus naturally encoded in SH/PFC models. Introducing a small perturbation in the periodicity of the order parameter and neglecting higher-order terms, the energy can be written as a quadratic form of the strain \cite{Elder2004,HeinonenPRE2014} resembling classical models of linear elasticity. Elastic constants are then related to the lattice symmetry described by $\psi$, with magnitudes dependent on the model parameters. Advanced models have also been proposed to account for elastic relaxation timescales \cite{Stefanovic2006,HeinonenPRE2014,Toth_2014,HeinonenPRL2016,SkogvollJMPS2022}.
Without approximations, the elastic energy obtained by perturbing the density field also comprises higher-order terms, including nonlinear and strain-gradient terms. While nonlinearities have been the object of dedicated analyses \cite{Huter2016,Huter2017,salvalaglio2022coarse}, strain-gradient terms have typically been neglected. However, numerical simulations exhibit features typical of GE. For instance, the stress field at dislocations in PFC simulations matches classical continuum elasticity in the far field but is nonsingular at the core\cite{SkaugenPRB2018,SalvalaglioNPJ2019,SalvalaglioJMPS2020,salvalaglio2022coarse}. Qualitatively, some form of regularization is indeed expected as the fields $\psi$ are smooth functions of the spatial coordinates due to the underlying free energy functionals. However, the emerging regularization explicitly resembles that obtained in GE theories \cite{Lazar2005}, for which preliminary evidence has been reported \cite{SalvalaglioJMPS2020,salvalaglio2022coarse}.

In this work, we study how GE is encoded in SH/PFC models. We bridge for the first time the field of nonlocal elasticity encoding microscopic effects (as described by GE) with the theory of pattern formation at the micro-to-meso-scale conveyed by SH/PFC models, which also serve as convenient frameworks for numerical simulations. The importance of establishing this connection is thus two-fold. (i) Current approaches in the field of GE typically assume some form of elastic energy; here, we show how GE free energies and constants entering this theory follow from microscopic models like SH and PFC, as well as related aspects, like the regularization of elastic fields at defects. (ii) By showing analytically and numerically that strain-gradient terms are inherently encoded in the SH free energy, we generally assess that approaches based on SH and PFC models retain these effects and discuss how GE emerges with concrete examples. Our work then enables advanced interpretation of simulation results. It paves the way for using SH and PFC models to study GE effects self-consistently, i.e., following assumptions on microscopic ordering in lattices or, more generally, patterns retaining some degree of order.

This paper is organized as follows. In Sect.~\ref{sec:GE}, we recall the basics of the GE theory, reporting the key information exploited in the current analysis. Readers familiar with this theory may skip this section and use it as a reference for the adopted notation. Similarly, in Sect.~\ref{sec:SHPFC}, we illustrate the SH free energy functional and recall the basics of SH and PFC models. 
From Sect.~\ref{sec:elas-sh}, we discuss novel aspects and results. As \textit{first main focus}, we discuss how GE naturally arises from the SH energy functional upon perturbation of the microscopic density (Sect.~\ref{sec:elasge}). We then analyze in detail the derived theory for stripe (Sect.~\ref{sec:stripe}) and crystalline (Sect.~\ref{sec:crystals}) phases, followed by a discussion of the results and their implications (Sect.~\ref{sec:discussion4}). While this analysis is obtained for a minimal formulation of the SH energy functional, extended parametrizations achieved via the definition of a two-point correlation function are also discussed (Sect.~\ref{sec:xpfcge}). As the \textit{second main focus}, we address in Sect.~\ref{sec:examples} the effective GE theory that transpires without approximations in numerical simulations. We study, in particular, the relevant case of the elastic fields at dislocations. After reporting the details concerning numerical simulations (Sect.~\ref{sec:num5}), we characterize the stress field and its regularization at the core of an edge dislocation via comparisons with analytic solutions from first and second strain-gradient elasticity and fitting of effective characteristic lengths (Sect.~\ref{sec:elas5}). Results including the dependence of the characteristic lengths on model parameters, like the quenching depth, and their implications, are then discussed (Sect.~\ref{sec:discussion5}). Conclusions are summarized in Sect.~\ref{sec:conclusions}.
Technical details concerning numerical methods, analytic expressions for the stress fields of dislocations in first and second strain-gradient elasticity, details of fitting procedures, and additional information, are reported in the Appendix. 

\section{Gradient Elasticity}
\label{sec:GE}
In \textit{first strain-gradient elasticity} (GE-1), the elastic energy density accounts for strain and its first derivatives. In the Toupin–Mindlin theory of anisotropic GE-1 \cite{toupin1962elastic,toupin1964theories}, following the convenient notation of Refs.~\cite{Lazar2018jmmp,Lazar2022_iso}, the energy density generally reads \footnote{We hereafter adopt Einstein summation convention.}
\begin{equation}\label{eq:eldensity}
    w(\boldsymbol{\varepsilon},\nabla \boldsymbol{\varepsilon})=\frac{1}{2}\mathbb{C}_{ijkl}\varepsilon_{ij}\varepsilon_{kl}+\frac{1}{2} \mathbb{D}_{ijmkln}\partial_m\varepsilon_{ij}\partial_n\varepsilon_{kl}+\mathbb{E}_{ijklm}\varepsilon_{ij}\partial_m\varepsilon_{kl},
\end{equation}
with $\boldsymbol\varepsilon=\frac{1}{2}(\nabla\mathbf{u} + \nabla\mathbf{u}^\intercal)$ the strain tensor, $\mathbf{u}$ the \textit{macroscopic} displacement field, and $\mathbb{C}$, $\mathbb{E}$, and $\mathbb{D}$, the constitutive four-, five- and six-rank tensors encoding elastic constants. This formulation simplifies the more general Mindlin's theory of elasticity in the presence of microstructures, which includes quantities at both macro- and micro-scale \cite{Mindlin1964,ASKES20111962}. This multiscale formulation, however, can hardly be exploited for practical purposes, and gradient elasticity theories based on the macroscopic displacement $\mathbf{u}$ are usually considered. In practice, the role played by microscopic deformation gradients is encoded by second gradients of the macroscopic displacement (Form I) or first gradients of the macroscopic strain (Form II), the latter leading to Eq.~\eqref{eq:eldensity}.
$\mathbb{C}$ and $\mathbb{D}$ possess both minor and major symmetries, while $\mathbb{E}$ has only minor symmetries. Illustrated for $\mathbb{C}$, these symmetries read component-wise
\begin{equation}
\mathbb{C}_{ijkl} =\mathbb{C}_{klij} \ \ \ \text{(major)},
    \qquad    \qquad \mathbb{C}_{ijkl}=\mathbb{C}_{jikl}=\mathbb{C}_{ijlk} \ \ \ \text{(minor)}.
\end{equation}
In the most general case, the constitutive tensors have 21, 108, and 171 independent components, respectively (in 3D), which may be derived from the energy density
\begin{equation}
    \mathbb{C}_{ijkl} = \frac{\partial^2w}{\partial \varepsilon_{ij} \partial \varepsilon_{kl}}, \qquad
    \mathbb{E}_{ijklm} = \frac{\partial^2w}{\partial \varepsilon_{ij} \partial (\partial_m \varepsilon_{kl})}, \qquad
    \mathbb{D}_{ijmkln} = \frac{\partial^2w}{\partial (\partial_m\varepsilon_{ij}) \partial (\partial_n \varepsilon_{kl})}.
\end{equation}
The quantities conjugate to the elastic strain and the gradient of the elastic strain are the Cauchy stress $\boldsymbol\sigma$ and the double stress $\boldsymbol\tau$
\begin{eqnarray}
    \sigma_{ij} &= \frac{\partial w}{\partial \varepsilon_{ij}} = \mathbb{C}_{ijkl}\varepsilon_{kl}+\mathbb{E}_{ijklm}(\partial_m\varepsilon_{kl}),
    \\
    \tau_{ijm} &= \frac{\partial w}{\partial (\partial_m\varepsilon_{ij})} = \mathbb{E}_{klijm}\varepsilon_{kl}+\mathbb{D}_{ijmkln}(\partial_n\varepsilon_{kl}).
\end{eqnarray}
In isotropic and, more generally, centrosymmetric crystals, $\mathbb{E}_{ijklm}=0$, resulting in the so-called Mindlin GE-1 for centrosymmetric material \cite{Mindlin1964,Mindlin1968}
\begin{equation}
    w(\boldsymbol{\varepsilon},\nabla \boldsymbol{\varepsilon})=\frac{1}{2}\mathbb{C}_{ijkl}\varepsilon_{ij}\varepsilon_{kl}+\frac{1}{2} \mathbb{D}_{ijmkln}\partial_m \varepsilon_{ij}\partial_n\varepsilon_{kl}.
\end{equation}
The mechanical equilibrium condition (without body forces) reads 
\begin{equation}
    \partial_j (\sigma_{ij}-\partial_m\tau_{ijm})=0,  
\end{equation}
which can be recast in a field equation for the displacement \cite{Lazar2018jmmp}:
\begin{equation}\label{eq:likdc}
\begin{split}
    L_{ik}u_k &= 0,
    \\
    L_{ik} &= L_{ik}^{\mathbb{C}}-L^{\mathbb{D}}_{ik} = \mathbb{C}_{ijkl}\partial_j\partial_l-\mathbb{D}_{ijmkln}\partial_j\partial_l\partial_m\partial_n.
\end{split}    
\end{equation}

In the following, Voigt notation is also used for the fourth-order tensor $\mathbb{C}$
\begin{equation}
11 \rightarrow 1,\ \ \ \
22 \rightarrow 2,\ \ \ \
33 \rightarrow 3,\ \ \ \
23=32 \rightarrow 4,\ \ \ \  
13=31 \rightarrow 5,\ \ \ \
12=21 \rightarrow 6,\ \ \ \
\end{equation}
with components denoted $C_{ij}$, and for the six-order tensor $\mathbb{D}$
\begin{equation}
\begin{array}{lllll}
    111 \rightarrow 1,\ \
    &221 \rightarrow 2,\ \
    &122 \rightarrow 3,\ \
    &331 \rightarrow 4,\ \
    &133 \rightarrow 5,\ \
    \\
    222 \rightarrow 6,\ \
    &332 \rightarrow 7,\ \
    &233 \rightarrow 8,\ \
    &112 \rightarrow 9,\ \
    &211 \rightarrow 10,\ \
    \\
    333 \rightarrow 11,\ \
    &113 \rightarrow 12,\ \
    &311 \rightarrow 13,\ \
    &223 \rightarrow 14,\ \
    &322 \rightarrow 15,\ \
    \\
    123 \rightarrow 16,\ \
    &132 \rightarrow 17,\ \
    &231 \rightarrow 18. & &
\end{array}
\end{equation}
with components denoted $D_{i,j}$.

\subsection{Isotropic materials}
\label{sec:GE1}

For isotropic materials ({iso}), the elastic tensors read
\begin{equation}\label{eq:Ciso}
\mathbb{C}_{ijkl}^{\rm iso}=\lambda\delta_{ij}\delta_{kl}+\mu(\delta_{ik}\delta_{jl}+\delta_{il}\delta_{jk}),
\end{equation}
with $\lambda$ and $\mu$ the Lam\'e constants, and
\begin{equation}\label{eq:ge_iso}
\begin{split}
    \mathbb{D}_{ijmkln}^{\rm iso} &= \frac{a_1}{2}(\delta_{ij}\delta_{km}\delta_{ln}+\delta_{ij}\delta_{kn}\delta_{lm}+\delta_{kl}\delta_{im}\delta_{jn} +\delta_{kl}\delta_{in}\delta_{jm}) 
    + 2a_2\delta_{ij}\delta_{kl}\delta_{mn} 
    \\
    &+\frac{a_3}{2}(\delta_{jk}\delta_{im}\delta_{ln}+\delta_{ik}\delta_{jm}\delta_{ln}+\delta_{il}\delta_{jm}\delta_{kn}+\delta_{jl}\delta_{im}\delta_{kn})
    + a_4(\delta_{il}\delta_{jk}\delta_{mn}+\delta_{ik}\delta_{jl}\delta_{mn}) 
    \\
    &+\frac{a_5}{2}(\delta_{jk}\delta_{in}\delta_{lm}+\delta_{ik}\delta_{jn}\delta_{lm}+\delta_{jl}\delta_{km}\delta_{in}+\delta_{il}\delta_{km}\delta_{jn}),
\end{split}
\end{equation}
with $a_i$ coefficients of the strain-gradient terms. We thus obtain the elastic energy density
\begin{equation}\label{eq:wiso}
\begin{split}
    w_{\rm iso}&=\frac{1}{2}\lambda\varepsilon_{ii}\varepsilon_{jj}+\mu\varepsilon_{ij}\varepsilon_{ij} 
    + a_1 (\partial_{j}\varepsilon_{ii})(\partial_{k}\varepsilon_{jk}) 
    + a_2 (\partial_{j}\varepsilon_{ii})(\partial_{j}\varepsilon_{kk})
    \\
    &+a_3 (\partial_{i}\varepsilon_{ij})(\partial_{k}\varepsilon_{jk})
    + a_4 (\partial_{k}\varepsilon_{ij})(\partial_{k}\varepsilon_{ij})
    + a_5 (\partial_{k}\varepsilon_{ij})(\partial_{i}\varepsilon_{jk}),
\end{split}
\end{equation}
and the operators
\begin{align} 
    L_{ik}^{\mathbb{C},\,{\rm iso}} &= (\lambda+2\mu)\partial_i\partial_k+\mu(\delta_{ik}\Delta-\partial_i\partial_k), \label{eq:likc}
    \\
    L_{ik}^{\mathbb{D},\,{\rm iso}} &= 2(a_1+a_2+a_3+a_4+a_5)\partial_i\partial_k\Delta  + \frac{1}{2}(a_3+2a_4+a_5)(\delta_{ik}\Delta - \partial_i\partial_k) \label{eq:likd}.   
\end{align}
Equations \eqref{eq:likc} and \eqref{eq:likd} correspond to the expressions in \eqref{eq:likdc} with (isotropic) elastic constants as in \eqref{eq:Ciso} \cite{Lazar2018jmmp,Lazar2022_iso}. Moreover, one may redefine $L_{ik}^{\rm iso}$ as
$$
L_{ik}^{\rm iso}=(\lambda+2\mu)[1-\ell_1^2\nabla^2]\partial_i\partial_k+\mu[1-\ell^2_2\nabla^2](\delta_{ik}\nabla^2 -\partial_i\partial_k),
$$
with $\nabla^2$ the Laplacian and the two emerging \textit{characteristic lengths}
\begin{align}
    \ell^2_1 &= \frac{2(a_1+a_2+a_3+a_4+a_5)}{\lambda+2\mu}, \\
    \ell^2_2 &= \frac{a_3+2a_4+a_5}{2\mu}.
\end{align}
$\ell_i$ can be considered measures of the material's internal characteristic lengths, such as the distance over which microstructural effects (like atomic lattice or grain structure) significantly influence the material's response to deformation. In essence, they provide a way to include the effect of the material's microstructure in continuum-scale models. $a_i$ may be expressed in terms of the components of $\mathbb{D}$, and computed by atomistic methods \cite{Lazar2022_iso}. In the following, they are directly extracted by comparison of the considered elastic energy densities with the form given in \eqref{eq:wiso}, see also Sect.~\ref{sec:elasge}.

\subsection{Centrosymmetric materials}

Centrosymmetric materials ({cs}) possess a lattice structure with an inversion center $\mathbf{r}_0$, i.e., points $\mathbf{r}_0\pm \mathbf{r}$ are equivalent. For these materials, elastic constants read
\begin{equation}\label{eq:Csqr}
\mathbb{C}_{ijkl}^{\rm cs}=C_{12}\delta_{ij}\delta_{kl}+C_{44}(\delta_{ik}\delta_{jl}+\delta_{il}\delta_{jk})+(C_{11}-C_{12}-2C_{44})\delta_{ijkl},
\end{equation}
with
\begin{equation}
\delta_{ijkl}=\left\{
\begin{array}{l}
    1, \quad\text{if } i=j=k=l, \\
    0, \quad\text{otherwise}.
\end{array}
\right.
\end{equation}
The constants entering the prefactor of $\delta_{ijkl}$ are typically used to quantify the anisotropy of a cubic material via, e.g., the so-called Zener ratio $z=2C_{44}/(C_{11}-C_{12})$. For isotropic materials, $z=1$ and Eq.~\eqref{eq:Csqr} reduces to Eq.~\eqref{eq:Ciso} as expected. The strain-gradient elasticity tensor can be written as \cite{Lazar2022_iso}
\begin{equation}
\begin{split}
    \mathbb{D}_{ijmkln}^{\rm cs}&=\frac{a_1}{2}(\delta_{ij}\delta_{km}\delta_{ln}+\delta_{ij}\delta_{kn}\delta_{lm}+\delta_{kl}\delta_{im}\delta_{jn} +\delta_{kl}\delta_{in}\delta_{jm})
    +2a_2\delta_{ij}\delta_{kl}\delta_{mn} 
    \\
    &+\frac{a_3}{2}(\delta_{jk}\delta_{im}\delta_{ln}+\delta_{ik}\delta_{jm}\delta_{ln}+\delta_{il}\delta_{jm}\delta_{kn}+\delta_{jl}\delta_{im}\delta_{kn})    +a_4(\delta_{il}\delta_{jk}\delta_{mn}+\delta_{ik}\delta_{jl}\delta_{mn}) 
    \\
    &+\frac{a_5}{2}(\delta_{jk}\delta_{in}\delta_{lm}+\delta_{ik}\delta_{jn}\delta_{lm}+\delta_{jl}\delta_{km}\delta_{in}+\delta_{il}\delta_{km}\delta_{jn}) 
    \\    &+a_6(\delta_{ik}\delta_{jlmn}+\delta_{il}\delta_{jkmn}+\delta_{jk}\delta_{ilmn}+\delta_{jl}\delta_{ikmn}) 
    \\    &+a_7(\delta_{km}\delta_{ijln}+\delta_{lm}\delta_{ijkn}+\delta_{in}\delta_{jklm}+\delta_{jn}\delta_{iklm}) 
    \\
    &+a_8\delta_{mn}\delta_{ijkl}
    +a_9(\delta_{ij}\delta_{klmn}+\delta_{kl}\delta_{ijmn}) 
    \\
    &+a_{10}(\delta_{im}\delta_{jkln}+\delta_{jm}\delta_{ikln}+\delta_{kn}\delta_{ijlm}+\delta_{ln}\delta_{ijkm})
    +a_{11}\delta_{ijklmn}.        
\end{split}
\end{equation}
With these elasic constants, the operator $L_{ik}^{\rm cs}$ for the generalized equations for displacements \eqref{eq:likdc} becomes \cite{Lazar2022_iso}: 
\begin{equation}
\begin{split}
    L_{ik}^{\rm cs} &= (C_{11}+2C_{44})[1-\ell_1^2\nabla^2]\partial_i\partial_k+C_{44}[1-\ell^2_2\nabla^2](\delta_{ik}\nabla^2 -\partial_i\partial_k)
    \\
    &+(C_{11}-C_{12}-2C_{44})[1-\ell^2_3\nabla^2]\delta_{ik}\partial_i\partial_k-(a_6\delta_{ik}\delta_{jlmn}+a_{11}\delta_{ijklmn})\partial_j\partial_l\partial_m\partial_n
    \\
    &-(a_6+a_7+a_9+a_{10})(\delta_{klmn}\partial_i+\delta_{ilmn}\partial_k)\partial_l\partial_m\partial_n.
\end{split}
\end{equation}
with characteristic lengths
\begin{equation}\label{eq:ge_lengths}
\begin{split}
    \ell^2_1 &= \frac{2(a_1+a_2+a_3+a_4+a_5)}{C_{12}+2C_{44}}, 
    \\
    \ell^2_2 &= \frac{a_3+2a_4+a_5}{2C_{44}}, 
    \\
    \ell^2_3 &= \frac{a_6+2a_7+a_8+2a_{10}}{C_{11}-C_{12}-2C_{44}},
    \end{split}
\end{equation}
In this case, terms explicitly depending on $a_i$ appear in $L_{ik}^{\rm cs}$, marking the intrinsic anisotropy of cubic crystals.

\subsection{Special second strain-gradient elasticity}
\label{sec:GE2}

In \textit{second strain-gradient elasticity} (GE-2), second derivatives of the strain field are retained in the elastic energy density, $w\equiv w(\varepsilon_{ij},\partial_k \varepsilon_{ij}, \partial_k\partial_\ell\varepsilon_{ij})$. For isotropic materials, a formulation relying on two characteristic lengths $\varrho_{1,2}$ may be considered. They enter the mechanical equilibrium condition as \cite{LAZAR20061787}
\begin{equation}\label{eq:mechge-2}
\begin{split}
\partial_j\left[ \sigma_{ij}-\partial_m\tau_{ijm}+\partial_n\partial_m\tau_{ijmn} \right] &= \partial_j\left[ (1-\varrho_1^2\nabla^2)(1-\varrho_2^2\nabla^2)\sigma_{ij}\right] = 0\\
&=\partial_j\left[ (1-\omega^2 \nabla^2+\gamma^4  \nabla^4)\sigma_{ij}\right] = 0,
\end{split}
\end{equation}
with $\tau_{ijmn}=\partial f/\partial(\partial_n \partial_m \varepsilon_{ij})$ the triple stress tensor and $\nabla^4$ the bi-Laplacian. Constants relate as $\omega^2=\varrho_1^2+\varrho_2^2$ and $\gamma^4=\varrho_1^2\varrho_2^2$. Note that to leading order ($\gamma=0$), this theory reduces to a single characteristic length $\omega$, i.e. to GE-1 with $\ell_1=\ell_2=\omega$.
This formulation has been exploited to obtain analytical solutions, e.g., for dislocation-induced deformations, which will be used in the following.

\section{Swift-Hohenberg and phase-field crystal models}
\label{sec:SHPFC}

\subsection{Free Energy for the microscopic density field}
The Swift-Hohenberg (SH) and phase-field crystal (PFC) models describe the nonconservative and conservative dynamics of a periodic order parameter $\psi \equiv \psi(\mathbf{r})$, respectively. The dynamical equation can be written in terms of the variation of a free energy functional $F$ that, in its simplest form, can be written \cite{Swift1977,Elder2002}
\begin{equation}\label{eq:adimPFC}
    F=\int_{\Omega} {\rm d}\mathbf{r} \bigg[\frac{1}{2} \psi \mathcal{L} \psi + \frac{1}{4}\psi^4  \bigg],
\end{equation}
with $\mathcal{L}=(q_0^2+\nabla^2)^2-\epsilon$.
This differential operator enforces a periodicity in agreement with the first peak in the structure factor (at $q_0$) for periodic patterns like stripes or crystalline phases. $\epsilon$ is a phenomenological temperature parameter. The so-called SH equation corresponds to the nonconservative \mbox{$L^2$-gradient} flow
\begin{equation}
\partial_t \psi = -\frac{\delta F}{\delta \psi} = [\epsilon- (q_0^2+\nabla^2)^2]\psi - \psi^3.
\end{equation}
The dynamics equation in the PFC model is instead given by the conservative \mbox{$H^{-1}$-gradient} flow
\begin{equation}\label{eq:dynPFC}
\partial_t \psi = \nabla^2 \frac{\delta F}{\delta \psi} = \nabla^2 \bigg\{ [(q_0^2+\nabla^2)^2-\epsilon]\psi + \psi^3 \bigg\}.
\end{equation}

Eq.~\eqref{eq:adimPFC} realizes the minimal free energy to model order-disorder phase transition as its minimizers are either constant or periodic fields. As such, its form emerges in different contexts. $F$ can be obtained as the nondimensional form of the free energy describing liquid-solid transition with the order parameter corresponding to the microscopic density field \cite{Elder2004}. It may also be derived---upon approximations---from the classical density functional theory \cite{Ramakrishnan1979,evans1979nature,singh1991density} as discussed, e.g., in Refs.~\cite{Elder2007,vanTeeffelen2009,Provatas2010,Archer2019}. In this approach, the differential operator can be obtained as an approximation of an $n$-point correlation function. Other approximations may also be considered, like the multimode SH energy functional \cite{MkhontaPRL2013}, effectively reproducing $M$ peaks at wavenumbers $q_m$ of the structure factor via a differential operator 
\begin{equation}
\mathcal{L}_M=\prod_{m=1}^{M}\left[(q_m^2+\nabla^2)^2+b_m\right],
\end{equation}
where the $b_m$ are additional constants controlling the stability of the corresponding modes. 
Without loss of generality, we consider here the multimode SH free-energy functional with the following generalized parameterization 
\begin{equation}\label{eq:PFC3}
F_\psi=\int_{\Omega} {\rm d}\mathbf{r} \bigg[ A \psi \mathcal{L}_M \psi + B \psi^2 + C\psi^3+D \psi^4 \bigg].
\end{equation}

\subsection{Amplitude expansion}
\label{sec:apfc}
The density $\psi$ can generally be expressed in terms of a small set of Fourier modes
\begin{equation}
\psi(\mathbf{r})=\psi_0(\mathbf{r})+\sum_{n=1}^N \eta_n(\mathbf{r}) {\rm e}^{{\rm i}\kv_n \cdot \mathbf{r}}+{\rm c.c.},
\label{eq:psipsi}
\end{equation}
with $\kv_n$ the reciprocal space vectors and c.c. denoting the complex conjugate. For a relaxed crystal, amplitudes are real functions $\eta_n=\varphi_n$ with $\varphi_i=\varphi_j$ if $|\mathbf{k}_i|=|\mathbf{k}_j|$. A deformed crystal can be described by Eq.~\eqref{eq:psipsi} with amplitudes being complex functions $\eta_n(\mathbf{r})=\varphi_n(\mathbf{r}) e^{-{\rm i} \kv_n \cdot \mathbf{u}(\mathbf{r})}$ \cite{Spatschek2010,ElderPRE2010,Huter2016}.
These complex amplitudes allow for the separation of different physical features \cite{HeinonenPRE2014}. For small deformations, the phase $\theta_n=-\kv_n \cdot \mathbf{u}$ encodes the deformation field, while $\varphi_n(\mathbf{r})$ accounts for diffusive phenomena, melting or solidification (e.g., a transition from finite $\varphi_n(\mathbf{r})$ to $\varphi_n(\mathbf{r})=0$ would be representative of an order-disorder / solid-liquid interface). 

In the amplitude expansion of the PFC model (APFC) \cite{salvalaglio2022coarse}, a free energy $F_\eta$ dependent on amplitudes can be derived from $F_\psi$. For a one-mode approximation of $\psi$ ($M=1$ in $\mathcal{L}_M$), this has been obtained via a renormalization group approach or by substituting \eqref{eq:psipsi} into the expression of the free energy functional and integrating over the unit cell \cite{Goldenfeld2005,AthreyaPRE2006,Yeon2010}. Here, with the shortest reciprocal-space vector set to $|\kv_n|=q_1=1$
and focusing on bulk systems where $\psi_0(\mathbf{r})$ can be assumed constant ($\psi_0(\mathbf{r})=\bar{\psi}$), we consider the amplitude approximation of the multimode SH functional \eqref{eq:PFC3} introduced in \cite{DeDonnoPRM2023}
\begin{equation}
F_\eta
=\int_{\Omega}\rmd\mathbf{r} \left[ \sum_{n=1}^N \left( A^\prime\Gamma_n|\mathcal{G}_{n}\eta_n|^2 \right) +g^{\rm s}(\{\eta_n\})  \right], 
\label{eq:APFC}
\end{equation}
with $\mathcal{G}_{n}=\nabla^2+2{\rm i}\kv_n\cdot\nabla$, $g^{\rm s}
    =\frac{B^\prime}{2}\zeta_2
    +\frac{C^\prime}{3}\zeta_3
    +\frac{D^\prime}{4}\zeta_4
    +E^\prime$, $\zeta_{p}$ complex polynomials of order $p$ in the amplitudes depending on the lattice symmetry (see Ref.~\cite{salvalaglio2022coarse}), and
\begin{equation}
\Gamma_n=\frac{\prod\limits_{\substack{
    m=1 \\
    q_m\neq|\kv_n|}}^M [(q_m^2-|\kv_n|^2)^2+b_m]}{\prod\limits_{m=2}^M [(q_m^2-q_1^2)^2+b_m]}
\end{equation}  
where the denominator is chosen to enforce $\Gamma_n=1$ for the first mode. Parameters of \eqref{eq:APFC} relate to the ones entering \eqref{eq:PFC3} as
 $A^\prime=A$, $B^\prime=B+2C\bar{\psi}+3D\bar{\psi}^2$, $C^\prime=C+3D\bar{\psi}$, $D^\prime=D$, and $E^\prime=(A/2)\bar{\psi}^2+(C/3)\bar{\psi}^3+(D/4)\bar{\psi}^4$. The dynamic equation for amplitudes approximating Eq.~\eqref{eq:dynPFC} is 
 \begin{equation}\label{eq:dynAPFC}
 \partial_t \eta_n = - |\kv_n|^2 \frac{\delta F_\eta}{\delta \eta_n^*}.
 \end{equation}

\section{Elasticity from the SH functional, constants, and characteristic lengths}
\label{sec:elas-sh}

\subsection{Elastic energy with strain-gradient terms}
\label{sec:elasge}

As the displacement field varies over large length scales, elasticity in the SH or PFC model can be described by looking at amplitudes $\{\eta_j \}$ \cite{HeinonenPRE2014,salvalaglio2022coarse}. In this context, the elastic energy density, namely the part of the energy depending on the displacement $\mathbf{u}$, is
\begin{equation}
\mathcal{E}=\int_\Omega {w}(\{\eta_n\})\ d\mathbf{r}=\int_\Omega\sum_n A\Gamma_n|\mathcal{G}_n \eta_n|^2 d\mathbf{r}.
\end{equation}
To proceed further, we consider two widely adopted simplifications, assuming constant amplitude moduli ($|\nabla\varphi_n|=0$) and neglecting nonlinear terms. With the former assumption, we obtain
\begin{equation}\label{eq:geta}
\begin{split}
    |\mathcal{G}_n \eta_n|^2 = &4\varphi_n^2(\kv_n)_i (\kv_n)_j (\kv_n)_k (\kv_n)_\ell \times
    \\
    & \big[\underbrace{(\partial_i u_j)(\partial_k u_\ell)-
    (\partial_i u_j)(\partial_k u_r)(\partial_\ell u_r)+
    \frac{1}{4}(\partial_i u_r)(\partial_j u_r)(\partial_k u_s)(\partial_\ell u_s)}_{\mathbf{U}^2}\big] 
    \\
    &+\varphi_n^2(\kv_n)_i (\kv_n)_j \underbrace{(\partial_{rr} u_i)(\partial_{ss}  u_j)}_{\rm \mathbf{G}^2},
    \end{split}
\end{equation}
with $\mathbf{U}$ a nonlinear strain tensor and $\mathbf{G}$ encoding \textit{gradient-elasticity} contributions. Due to the nature of the amplitude functions, this simplifying assumption is exact in relaxed bulk, where the amplitudes are real and constant. However, it is an approximation at defects and interfaces, where the amplitude moduli vary measurably. 
Next, neglecting higher-order nonlinear terms, we have 
\begin{equation}\label{eq:fel}
     w = A' \sum_{n=1}^N \Gamma_n \varphi_n^2 \left\{ 4[(\kv_n)_i(\kv_n)_j(\partial_i u_j)]^2+[(\kv_n)_i\nabla^2 u_i]^2\right\},
\end{equation}
with the first and second terms in the sum corresponding to linear and strain-gradient elasticity terms, respectively. Due to the presence of higher-order displacement derivatives, it is important to check that this approximated expression of the elastic energy density complies with rotational invariance. Indeed, while the free energy of the system \eqref{eq:APFC} is rotationally invariant by construction \cite{salvalaglio2022coarse}, and therefore so is the elastic energy density implicitly encoded therein, the approximations made to reach \eqref{eq:fel} may affect that property. This is, however, not the case, given the form of the double-derivative terms. Explicitly, if the space coordinates undergo a rigid rotation with the associated matrix $\mathbf{R}$, the term $(\kv_n)_i\nabla^2 u_i$ transforms as:
\begin{equation}
    R_{ij}(\kv_n)_j\nabla^2(R_{ik}u_k) = \delta_{jk}(\kv_n)_j\nabla^2 u_k = (\kv_n)_j\nabla^2 u_j,
\end{equation}
i.e., it is rotationally invariant. We refer to Refs.~\cite{munch2018,Brannon2018} for more detailed discussions. In the following, we look at the expressions of the elastic energy density obtained for specific patterns and lattice symmetries.

\subsection{Stripe Phase}
\label{sec:stripe}

Stripe arrangements, e.g., in smectic phases \cite{Elder1992, ElderPRA1992, Praetorius2018, HuangCP2022}, are a class of patterns described by SH and PFC models \cite{Swift1977,Elder2002,Thiele2013}. The corresponding minimizer of the SH energy functional is well represented by waves with a single reciprocal lattice vector $\kv$, namely with $\psi$ described as in Eq.~\eqref{eq:psipsi} with a single term in the sum. Assuming an arbitrarily oriented wave vector $\kv=k_0(\cos(\phi),\sin(\phi))$, Eq.~\eqref{eq:fel} reduces to:
\begin{equation}
\begin{split}
    \frac{w_{\rm str}}{K_{\rm str}} =& 
    4 k_0^2(\cos^4(\phi) \varepsilon_{xx}^2 
    + \sin^4(\phi) \varepsilon_{yy}^2 
    + 2\cos^2(\phi)\sin^2(\phi) \varepsilon_{xx}\varepsilon_{yy} 
    + 4\cos^2(\phi)\sin^2(\phi) \varepsilon_{xy}^2) \\
    & + (\cos(\phi)\nabla^2 u_x + \sin(\phi)\nabla^2 u_y)^2,
\end{split}
\end{equation}
with $K_{\rm str} = A'\varphi^2 k_0^2$. This leads to anisotropic elastic constants
\begin{equation}
{C_{\rm str}}=8A'\varphi^2 k_0^4
\begin{pmatrix}
    \cos^4(\phi) & \cos^2(\phi)\sin^2(\phi) & \cos^3(\phi)\sin(\phi)
    \\
    * & \sin^4(\phi) & \cos(\phi)\sin^3(\phi)
    \\
    * & * & \cos^2(\phi)\sin^2(\phi)
\end{pmatrix}.
\end{equation}
Due to major and minor symmetries, there are 21 independent strain-gradient constants in 2D, reading (with $k_0=1$)
\begin{equation}
\begin{split}
    2\cos^2(\phi)&=D_{1,1}=D_{1,3}=D_{2,2}=D_{3,3}=-D_{1,2}=-D_{2,3},\\
    2\sin^2(\phi)&=D_{6,6}=D_{9,9}=D_{10,10}=D_{6,10}=-D_{6,9}=-D_{9,10},\\
    2\cos(\phi)\sin(\phi)&=D_{1,6}=D_{1,10}=D_{2,9}=D_{3,6}=D_{3,10}\\
    &=-D_{1,9}=-D_{3,9}=-D_{2,6}=-D_{2,10},
\end{split}
\end{equation}
Notice that the strain-gradient constants cannot be recast into the characteristic lengths $\ell_i$. Stripe phases exhibit significant anisotropy in diffusive processes, such as marked directionality during growth \cite{Thiele2013,Holl2021,Skogvoll_2022}. Here, we obtain that this phase also exhibits elastic anisotropy for linear and strain-gradient elasticity contributions.

\subsection{Crystalline phases}
\label{sec:crystals}

For crystalline phases, we find that Eq.~\eqref{eq:fel} may be rewritten as
\begin{equation}\label{eq:sg_displ}
    A' \sum_{n=1}^{N} \Gamma_n \varphi^2_n [(\kv_n)_i\nabla^2 u_i]^2 = K \sum_i(\nabla^2 u_i)^2,\end{equation}
given that, for each mode $m$ (i.e., for each family such that $|\mathbf{k}_n|=q_m$), the symmetry of the corresponding reciprocal lattice vectors (see also below explicit examples) imposes
\begin{equation}
\begin{split}
    \sum_{\substack{n=1 \\ q_m=|\kv_n|}}^N (\kv_n)_i^2 = Q_m, \qquad \text{and} \qquad
    \sum_{\substack{n=1 \\ q_m=|\kv_n|}}^N (\kv_n)_i(\kv_n)_j = 0,\quad \forall i \neq j,
\end{split}
\end{equation}
where $K$ and $Q_m$ depend on the specific lattice symmetry and the number of modes considered. 
To determine GE constants and characteristic lengths, we then need to express the quantity \eqref{eq:sg_displ} in terms of strain derivatives. In 2D systems, we have that
\begin{equation}\label{eq:GtoStrain}
\begin{split}
    (\nabla^2 u_x)^2+ (\nabla^2 u_y)^2 =& (\partial_{x}\varepsilon_{xx}+\partial_{yy}^2u_{x})^2+(\partial_{y}\varepsilon_{yy}+\partial_{xx}^2u_{y})^2
    \\
    =&[(\partial_{x}\varepsilon_{xx})^2
    +(\partial_{y}\varepsilon_{yy})^2]
    +4[\partial_{y}\varepsilon_{xy}\partial_{x}\varepsilon_{xx}+
    \partial_{x}\varepsilon_{xy}
    \partial_{y}\varepsilon_{yy}]
    \\
    &-2[\partial_{x}\varepsilon_{yy}\partial_{x}\varepsilon_{xx}+
    \partial_{y}\varepsilon_{xx}
    \partial_{y}\varepsilon_{yy}] 
    +[(\partial_{x}\varepsilon_{yy})^2
    +(\partial_{y}\varepsilon_{xx})^2]
    \\
    &+4[(\partial_{x}\varepsilon_{xy})^2+
    (\partial_{y}\varepsilon_{xy})^2]
    -4[\partial_{y}\varepsilon_{xy}
    \partial_{x}\varepsilon_{yy}
    +\partial_{x}\varepsilon_{xy}
    \partial_{y}\varepsilon_{xx}],
    \end{split}
\end{equation}
where we exploited the identity $\partial_{ii}^2u_j = \partial_i (2 \varepsilon_{ij}-\partial_j u_i) = 2\partial_i \varepsilon_{ij}-\partial_j \varepsilon_{ii}$. We may then compare terms of $w$ from Eq.~\eqref{eq:fel} to the general form of the elastic energy for isotropic materials \eqref{eq:ge_iso}:
\begin{equation}\label{eq:syst_ai}
    \left\{
    \begin{aligned}
    a_1+a_2+a_3+a_4+a_5 &= K  && {\rm from} \ \  (\partial_{x}\varepsilon_{xx})^2,
    \\
    a_1+2a_3 &= 4K  && {\rm from} \ \  (\partial_{y}\varepsilon_{xy}\partial_{x}\varepsilon_{xx}+\partial_{x}\varepsilon_{xy}\partial_{y}\varepsilon_{yy}),
    \\
    a_1+2a_2 &= -2K && {\rm from} \ \ (\partial_{x}\varepsilon_{yy}\partial_{x}\varepsilon_{xx}+\partial_{y}\varepsilon_{xx} \partial_{y}\varepsilon_{yy}),
    \\
    a_2+a_4 &= K && {\rm from} \ \ (\partial_{x}\varepsilon_{yy})^2+(\partial_{y}\varepsilon_{xx})^2,
    \\
    a_3+2a_4+a_5 &= 4K && {\rm from} \ \ (\partial_{x}\varepsilon_{xy})^2+(\partial_{y}\varepsilon_{xy})^2,
    \\
    a_1+2a_5 &= -4K && {\rm from} \ \ (\partial_{x}\varepsilon_{yy}\partial_{y}\varepsilon_{xy}+\partial_{y}\varepsilon_{xx}\partial_{x}\varepsilon_{xy}).
    \end{aligned}
    \right.
\end{equation}
By solving this system of (dependent) equations for $a_i$ we obtain
\begin{equation}
2a_2=-2K - a_1, \qquad 2a_3 = 4 K - a_1, \qquad 2a_4 =  K + a_1, \qquad  
 2a_5 = -4 K - a_1.
\end{equation}
For three-dimensional systems, the expressions of the coefficients $a_i$ are analogous. This follows by considering the additional term $(\nabla^2 u_z)^2$ on the left-hand side of Eq.~\eqref{eq:GtoStrain} and proper extension of the right-hand side. The material characteristic length scales, as can be directly deduced from the first and fifth equations in \eqref{eq:syst_ai}, are
\begin{equation}\label{eq:lengths}
\begin{split}
    \ell_1 = \sqrt{\frac{2K}{\lambda+2\mu}}, \qquad
    \ell_2 = \sqrt{\frac{2K}{\mu}},
\end{split}
\end{equation}
with $\ell_3=0$. This means that under the assumptions considered in this section (see Eqs.~\eqref{eq:geta} and \eqref{eq:fel}), the minimal SH energy functional encodes isotropic GE-1. We note that while no centrosymmetric crystal can verify the conditions for isotropic strain-gradient elasticity (due to the Hermann theorem \cite{Hermann1934}), for cubic materials that are nearly isotropic with respect to the constitutive four-rank tensor $\mathbb{C}$, a Voigt-type averaging procedure may be employed for the sixth-rank constitutive tensor $\mathbb{D}$ to compute effective isotropic GE constants \cite{Lazar2022_iso}.   

We consider in the following different crystalline phases: triangular, square, bcc, and fcc. For some representative cases, we analyze different numbers of modes. 
We report the elastic energy density for the different symmetries/approximations (set by specific choices for $\kv_n$ with $\min_{n=1}^N|\kv_n|=1$), the corresponding elastic constants, and GE characteristic lengths for small deformations and neglecting nonlinearities (see Table~\ref{tab:ge_lengths} and Fig.~\ref{fig:lengths}). The scaling factor $k_0$ is introduced such that $a_0=2\pi/k_0$ is the interatomic distance.

\subsubsection*{Triangular phase, one-mode approximation (tri-1):} 

\[
\kv_1=k_0(-1,-1/\sqrt{3}),\qquad  \kv_2=k_0(0,2/\sqrt{3}), \qquad \kv_3=k_0(1,-1/\sqrt{3}). 
\]
with $k_0=\sqrt{3}/2$. Elastic energy density:
\begin{equation}
    \frac{w_{\rm tri-1}}{K_{\rm tri-1}}= 4k_0^2(\varepsilon^2_{xx}+\varepsilon_{yy}^2)+\frac{8}{3}k_0^2\varepsilon_{xx}\varepsilon_{yy}+\frac{16}{3}k_0^2\varepsilon_{xy}^2 + (\nabla^2 u_x)^2+(\nabla^2 u_y)^2, 
\end{equation}
with $K_{\rm tri-1}=2A'\varphi_1^2 k_0^2$.

\subsubsection*{Triangular phase, two-mode approximation (tri-2):} 
\begin{gather*}
\kv_1=k_0(-1,-1/\sqrt{3}),\qquad  \kv_2=k_0(0,2/\sqrt{3}), \qquad \kv_3=k_0(1,-1/\sqrt{3}), \\
\kv_4=k_0(-1,-\sqrt{3}),\quad 
\kv_5=k_0(-1,\sqrt{3}),\quad 
\kv_6=k_0(2,0), 
\end{gather*}
with $k_0=\sqrt{3}/2$. Elastic energy density:
\begin{equation}
\frac{w_{\rm tri-2}}{K_{\rm tri-2}}=4\alpha^2 k_0^2(\varepsilon_{xx}^2+\varepsilon_{yy}^2) + \frac{8}{3}\alpha^2 k_0^2\varepsilon_{xx}\varepsilon_{yy}+\frac{16}{3}\alpha^2 k_0^2\varepsilon_{xy}^2
+(\nabla^2 u_x)^2+(\nabla^2 u_y)^2,
\end{equation}
with $K_{\rm tri-2}=2 A'(\varphi_1^2+3\varphi_2^2)k_0^2$ and $\alpha^2=(\varphi_1^2+9\varphi_2^2)/(\varphi_1^2+3\varphi_2^2)$.

\subsubsection*{Square lattice, two-mode approximation (sq-2):} 
\begin{gather*}
\kv_1=k_0(1,0),\quad \kv_2=k_0(0,1),\quad \kv_3=k_0(1,1),\quad \kv_4=k_0(1,-1),
\end{gather*}
with $k_0=1$. Elastic energy density:
\begin{equation}
\frac{w_{\rm sq-2}}{K_{\rm sq-2}} = 4k_0^2(\varepsilon_{xx}^2+\varepsilon_{yy}^2)+16\alpha^2 k_0^2\varepsilon_{xx}\varepsilon_{yy}+32\alpha^2 k_0^2\varepsilon_{xy}^2 +(\nabla^2 u_x)^2+(\nabla^2 u_y)^2,
\end{equation}
with $K_{\rm sq-2} = A'(\varphi_1^2 + 2\varphi_2^2)k_0^2$ and $\alpha^2=\varphi_2^2/(\varphi_1^2+2\varphi_2^2)$.

\subsubsection*{BCC lattice, one-mode approximation (bcc-1):} 
\begin{gather*}
\kv_1=k_0\sqrt{3}/2(0,1,1),\quad
\kv_2=k_0\sqrt{3}/2(1,0,1),\quad
\kv_3=k_0\sqrt{3}/2(1,1,0),\\
\kv_4=k_0\sqrt{3}/2(0,1,-1),\quad
\kv_5=k_0\sqrt{3}/2(1,-1,0),\quad
\kv_6=k_0\sqrt{3}/2(-1,0,1),
\end{gather*}
with $k_0=\sqrt{2/3}$. Elastic energy density:
\begin{equation}
\begin{split}
    \frac{w_{\rm bcc-1}}{K_{\rm bcc-1}} &= 3k_0^2(\varepsilon_{xx}^2+\varepsilon_{yy}^2+\varepsilon_{zz}^2) 
    + 3k_0^2(\varepsilon_{xx}\varepsilon_{yy}+\varepsilon_{xx}\varepsilon_{zz}+\varepsilon_{yy}\varepsilon_{zz})  \\
    &+6k_0^2(\varepsilon_{xy}^2+\varepsilon_{xz}^2+\varepsilon_{yz}^2)
    + (\nabla^2 u_x)^2 + (\nabla^2 u_y)^2 + (\nabla^2 u_z)^2,
\end{split}
\end{equation}
with $K_{\rm bcc-1}=3A' k_0^2\varphi_1^2$. 

\subsubsection*{BCC lattice, two-mode approximation (bcc-2):} 
\begin{gather*}
\kv_1=k_0\sqrt{3}/2(0,1,1),\quad
\kv_2=k_0\sqrt{3}/2(1,0,1),\quad
\kv_3=k_0\sqrt{3}/2(1,1,0),\\
\kv_4=k_0\sqrt{3}/2(0,1,-1),\quad
\kv_5=k_0\sqrt{3}/2(1,-1,0),\quad
\kv_6=k_0\sqrt{3}/2(-1,0,1),\\
\kv_7=k_0\sqrt{3}(1,0,0),\quad
\kv_8=k_0\sqrt{3}(0,1,0),\quad
\kv_9=k_0\sqrt{3}(0,0,1), 
\end{gather*}
with $k_0=\sqrt{2/3}$. Elastic energy density:
\begin{equation}
\begin{split}
    \frac{w_{\rm bcc-2}}{K_{\rm bcc-2}} &= 3\frac{\varphi_1^2+4\varphi_2^2}{\varphi_1^2+\varphi_2^2}k_0^2(\varepsilon_{xx}^2+\varepsilon_{yy}^2+\varepsilon_{zz}^2) 
    + 3\alpha^2 k_0^2 (\varepsilon_{xx}\varepsilon_{yy}+\varepsilon_{xx}\varepsilon_{zz}+\varepsilon_{yy}\varepsilon_{zz}) 
    \\
    &+ 6\alpha^2 k_0^2 (\varepsilon_{xy}^2+\varepsilon_{xz}^2+\varepsilon_{yz}^2) 
    + (\nabla^2 u_x)^2 + (\nabla^2 u_y)^2 + (\nabla^2 u_z)^2,
\end{split}
\end{equation}
with $K_{\rm bcc-2}=3A' k_0^2(\varphi_1^2+\varphi_2^2)$ and $\alpha^2=\varphi_1^2/(\varphi_1^2+\varphi_2^2)$.

\subsubsection*{FCC lattice, two-mode approximation (fcc-2):} 
\begin{gather*}
\kv_1=k_0/\sqrt{2}(-1,1,1),\quad
\kv_2=k_0/\sqrt{2}(1,-1,1), \\
\kv_3=k_0/\sqrt{2}(1,1,-1),\quad
\kv_4=k_0/\sqrt{2}(-1,-1,-1), \\
\kv_5=k_0\sqrt{2}(1,0,0),\quad
\kv_6=k_0\sqrt{2}(0,1,0),\quad 
\kv_7=k_0\sqrt{2}(0,0,1),\quad
 \end{gather*}
with $k_0=\sqrt{2/3}$. Elastic energy density:
\begin{equation}
\begin{split}
    \frac{w_{\rm fcc-2}}{K_{\rm fcc-2}} &= 2\frac{\varphi_1^2+4\varphi_2^2}{\varphi_1^2+\varphi_2^2}k_0^2(\varepsilon_{xx}^2+\varepsilon_{yy}^2
    +\varepsilon_{zz}^2) 
    + 4\alpha^2k_0^2(\varepsilon_{xx}\varepsilon_{yy}+\varepsilon_{xx}\varepsilon_{zz}+\varepsilon_{yy}\varepsilon_{zz}) 
    \\
    & + 8\alpha^2 k_0^2(\varepsilon_{xy}^2+\varepsilon_{xz}^2+\varepsilon_{yz}^2) 
    + (\nabla^2 u_x)^2 + (\nabla^2 u_y)^2 + (\nabla^2 u_z)^2,
\end{split}
\end{equation}
with $K_{\rm fcc-2}=2A' k_0^2(\varphi_1^2+\varphi_2^2)$ and $\alpha^2=\varphi_1^2/(\varphi_1^2+\varphi_2^2)$.

\vspace{10pt}
\noindent The elastic constants and GE characteristic lengths obtained via Eq.~\eqref{eq:ge_lengths} from the elastic energy densities above are reported in \autoref{tab:ge_lengths}. Note that the constraint $C_{12}=C_{44}$ holds for all lattice symmetries. Interestingly, this corresponds to the Cauchy relation derived in the lattice-theoretical description of constitutive tensors for central force interaction between atoms \cite{Born1955,Lazar2022_iso}.
\begin{table}[ht]
\renewcommand{\arraystretch}{1.5}
\centering
\begin{tabular}{@{}cccccc@{}}
\toprule
\ Symmetry \ & 
\ $C_{11}$ ($A'k_0^4$)   \          & 
\ $C_{12}=C_{44}$ ($A'k_0^4$) \    &
\ $\ell_1 \ (a_0)$ \ &
\ $\ell_2 \ (a_0)$ \ &
\ $\alpha^2$ \ \\ \midrule
tri-1  & 
$16\varphi_1^2$  &        
$\frac{16}{3}\varphi_1^2$  &
$\frac{1}{4\pi}$ &
$\frac{\sqrt{3}}{4\pi}$  &
-- \\
tri-2    & 
$16(\varphi_1^2+9\varphi_2^2)$ & 
$\frac{16}{3}(\varphi_1^2+9\varphi_2^2)$ &
$\frac{1}{4\pi}\frac{1}{\alpha}$ &
$\frac{\sqrt{3}}{4\pi}\frac{1}{\alpha}$ &
$\frac{\varphi_1^2+9\varphi_2^2}{\varphi_1^2+3\varphi_2^2}$ \\
sq-2     &   
$8(\varphi_1^2+2\varphi_2^2)$ &  
$16\varphi_2^2$ &        
$\frac{1}{4\pi\sqrt{6}}\frac{1}{\alpha} $ & 
$\frac{1}{4\pi\sqrt{2}}\frac{1}{\alpha} $ &
$\frac{\varphi_2^2}{\varphi_1^2+2\varphi_2^2}$ \\
bcc-1    &     
$18\varphi_1^2$ &   
$9\varphi_1^2$ & 
$\frac{1}{3\pi\sqrt{2}}$ &
$\frac{1}{\pi\sqrt{6}}$ &
-- \\
bcc-2    &
$18(\varphi_1^2+4\varphi_2^2)$&
$9\varphi_1^2$&         
$\frac{1}{3\pi\sqrt{2}}\frac{1}{\alpha}$ &
$\frac{1}{\pi\sqrt{6}}\frac{1}{\alpha}$ &
$\frac{\varphi_1^2}{\varphi_1^2+\varphi_2^2}$ \\
fcc-2    &
$8(\varphi_1^2+4\varphi_2^2)$ &      
$8\varphi_1^2$ &
$\frac{1}{2\pi\sqrt{6}}\frac{1}{\alpha}$ &   
$\frac{1}{2\pi\sqrt{2}}\frac{1}{\alpha}$  &
$\frac{\varphi_1^2}{\varphi_1^2+\varphi_2^2}$ \\
\bottomrule
\end{tabular}
\caption{Elastic constants and characteristic lengths for the different lattice symmetries and approximations of the microscopic density with a different number of modes for triangular and bcc lattices.
}
\label{tab:ge_lengths}
\end{table}

\subsection{Discussion of the results}
\label{sec:discussion4}

\begin{figure}
    \centering
    \includegraphics[width=1.0\textwidth]{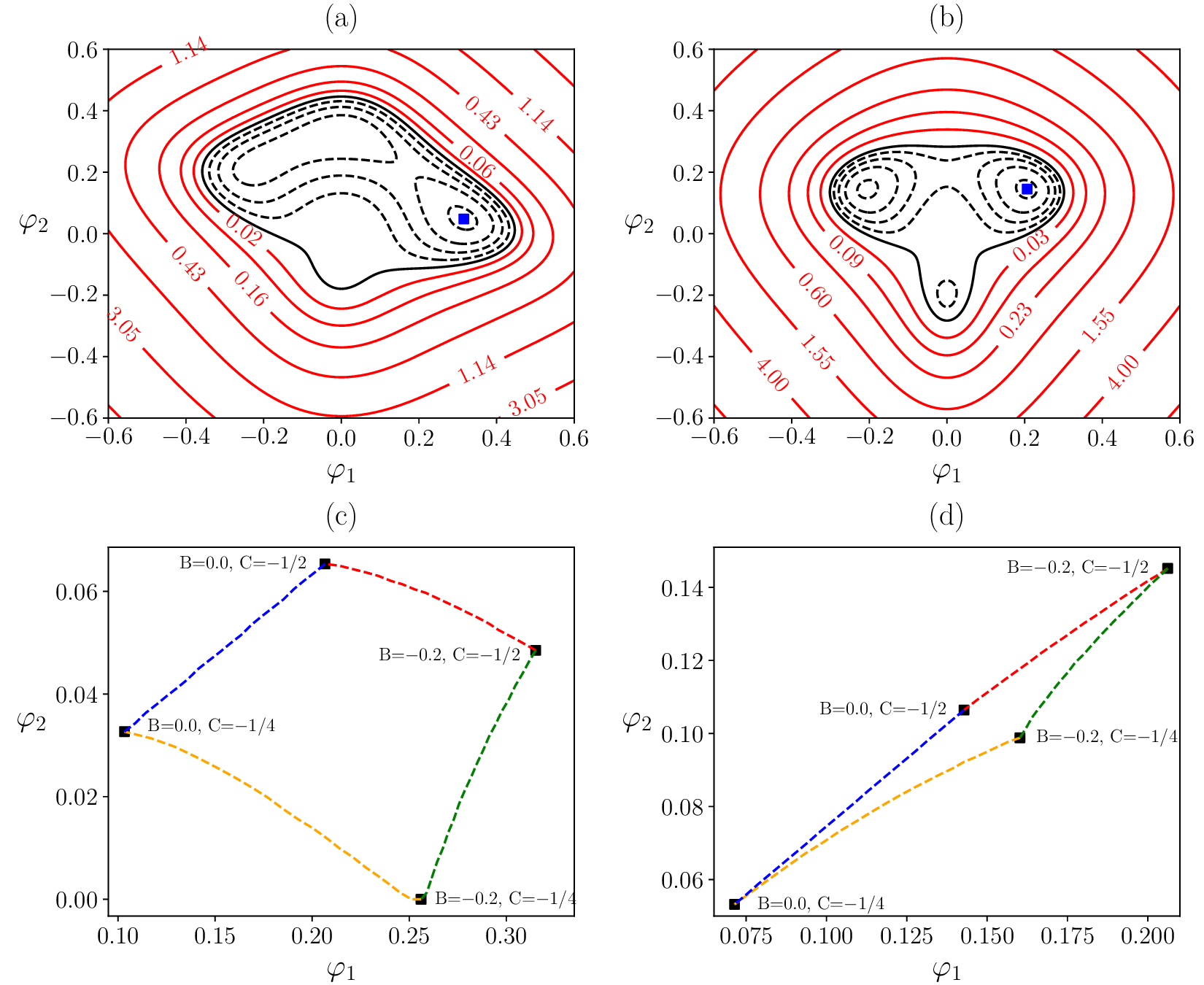}
    \caption{Bulk free energy density $f(\varphi_1,\varphi_2)$ for (a) triangular (2 modes)  and (b) fcc (2 modes) for $B=-0.2$, $C=-1/2$, and $D=1/3$. Solid red lines and dashed black lines are positive and negative values, respectively (values on contour lines are multiplied by $10^2$). The blue square indicates the global minimum. The variation of the amplitudes $\varphi_{1,2}$ minimizing the energy by varying parameters $B$ and $C$ (with $D=1/3$) is also illustrated for (c) triangular (2 modes) and (d) fcc (2 modes). Four representative configurations are illustrated (black squares). Dashed lines are obtained by varying the parameter that changes in the squares they connect.}
    \label{fig:phi1phi2}
\end{figure}

From the analysis reported above for crystalline phases, we find that $\ell_2/\ell_1=\sqrt{3}$ as a result of the general form of the (isotropic) strain-gradient terms; see Eq.~\eqref{eq:sg_displ}. Typical values found or adopted in the literature for this ratio are close to one \cite{ZHANG20062304,Lazar2023}.
For a one-mode approximation of $\psi$ (one length of $\kv$ considered), length scales are constants. For approximations of $\psi$ considering more than one mode, characteristic lengths depend on $\varphi_i$. Figure~\ref{fig:phi1phi2} illustrates the bulk free energy density for the \textit{tri-2} (Fig.~\ref{fig:phi1phi2}a) and \textit{fcc-2} (Fig.~\ref{fig:phi1phi2}b) case as a function of $\varphi_1$ and $\varphi_2$ for selected parameters. The global minima of these free energy densities (blue squares) correspond to the equilibrium phase. Values of $\varphi_1$ and $\varphi_2$ vary with the free energy parameters as dictated by the energy landscape, so they are not free parameters entering $\alpha$. An illustration of their variation is reported in Fig.~\ref{fig:phi1phi2}c (\textit{tri-2}) and \ref{fig:phi1phi2}d (\textit{fcc-2}).

\begin{figure}
    \centering
    \includegraphics[width=0.92\textwidth]{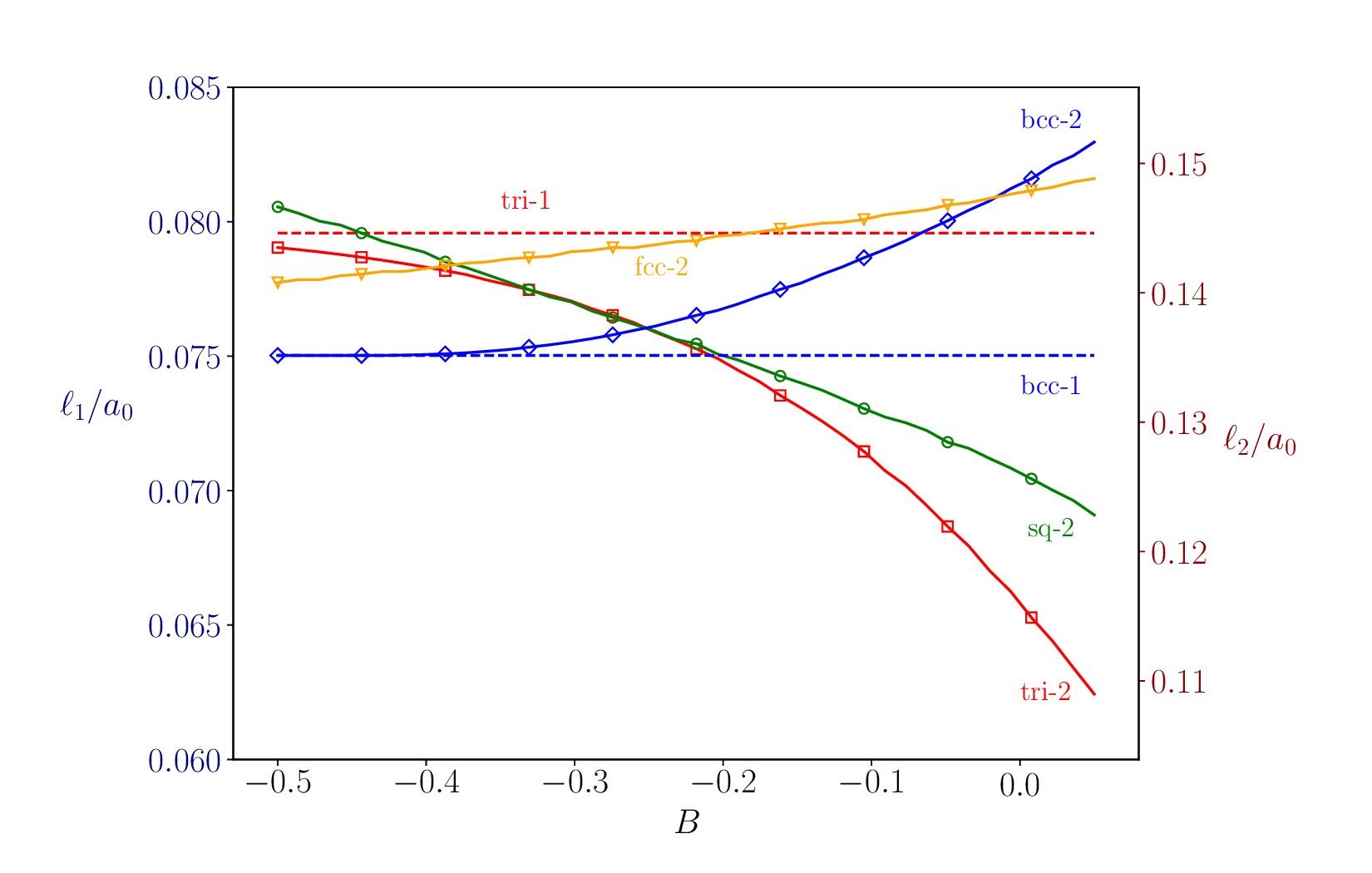}
    \caption{Characteristic lengths $\ell_1$ and $\ell_2$ by varying $B$ with $C=-1/2$ and $D=1/3$ as from Table \ref{tab:ge_lengths}. Dashed lines correspond to the length scales for one-mode approximations, which are only available for triangular and bcc symmetry.}
    \label{fig:lengths}
\end{figure}

Figure~\ref{fig:lengths} reports the values of $\ell_{1,2}$ by varying $B$ in the free energy for a selected parameter range where all the symmetries can be explored for $B<B_c$ and $B_c$ corresponding to the order-disorder critical point \cite{salvalaglio2022coarse}. Here, the effect of changes in $\varphi_i$ is then quantified for a specific choice of parameters. We note that a limited change is observed among all the explored ranges ($\lesssim 10\%$). The lattice symmetry affects the values of the GE length scales. At small enough $B$ (low ``temperature''), the number of modes considered does not impact the characteristic lengths, whereas a clear deviation is observed at larger $B$ (Fig.~\ref{fig:lengths}). 
In the context of APFC, these approximations realize different models as the choice of modes dictates the variables to solve for in the system (the amplitudes $\eta_j$). Instead considering a larger number of modes for the PFC and SH models results in a better estimation of their elastic properties.

The values of $\ell_{1,2}$ in Fig.~\ref{fig:lengths} are reported in terms of the ratio with the spatial length scale entering the free energy, namely the lattice parameter $a_0=2\pi/k_0$, with $k_0$ depending on the lattice symmetry (see specific values together with the definition of reciprocal space vectors in Sect. \ref{sec:crystals}). They all correspond to fractions of the lattice spacing. Therefore, they are in the order of $\sim 10^{-10}$m (\r{A}) when looking at physical units of common materials, in line with GE literature \cite{shodja2013ab,Lazar2023}. From a quantitative point of view, as we may expect considering the coarse-grained nature of SH and PFC models, we note that the ratio of the GE length scales with the lattice parameters does not precisely match typical values computed with \textit{ab initio} approaches \cite{shodja2013ab}. Suitable extensions by considering more advanced formulations \cite{Greenwood2010,Jaatinen2009} are outlined in the following section.

An important aspect emerging from this analysis is that the GE length scales ($\sim$ 1\r{A}) are typically smaller than the coarse-graining length that should be considered in PFC-like approaches to work with continuous elastic fields (filtering out microscopic fluctuation) or that underlies the derivation of the equations for APFC approaches. Therefore, while GE is self-consistently encoded, specific features emerging at those length scales are expected to be coarse-grained. This aspect is discussed further below with the aid of numerical examples.

\subsection{Characteristic lengths from structural PFC: triangular symmetry}
\label{sec:xpfcge}

Energy functionals generalizing Eq.~\eqref{eq:PFC3}, or Eq.~\eqref{eq:APFC} for amplitude formulations, can be devised by considering correlation function instead of the differential operator \cite{Greenwood2010,oforiopokuPRB20213}.
Such an approach proved powerful in devising parametrizations to model specific materials, e.g., in Ref.~\cite{Jaatinen2009}. Here, we outline the possibility of parametrizing the GE length scales by designing the correlation function. For a generic two-point correlation function $\corr$, the elastic energy takes the form:
\begin{equation}
    \mathcal{F}_{\rm el} = \mathcal{F}[\psi(\mathbf{r}+\mathbf{u}(\mathbf{r}))] - \mathcal{F}[\psi(\mathbf{r})],
\end{equation}
where
\begin{equation}
    \mathcal{F}[\psi(\mathbf{r})] = \int \mathrm{d}\mathbf{r}\, \bigg\{\psi(\mathbf{r}) \int \mathrm{d}\mathbf{r}^\prime \bigg[ \corr (\mathbf{r}-\mathbf{r}^\prime) \psi(\mathbf{r}^\prime) \bigg] \bigg\}.
\end{equation}
To allow for the analytical derivation of the elastic and gradient-elastic constants, we consider a polynomial expansion of $\corr$ in the reciprocal space---corresponding to the Dirac delta function and its derivatives in real space---so that the convolution can be carried out explicitly, leading to differentiation of $\mathbf{u}(\mathbf{r})$. We consider a polynomial expansion of the Fourier transform of $C_2$ reading
\begin{equation}\label{eq:exp_corrfun}
    \widehat{\corr}(|\kv|)=\sum_{n=0}^{N}c_{2n}|\kv|^{2n},
\end{equation}
where we only consider even terms due to symmetry considerations. For an $N$-term expansion, the highest order is thus $2N$.

We consider a setting that introduces additional parameters while reproducing the features of GE obtained by the minimal SH energy functional, focusing on the 1-mode triangular symmetry for illustration purposes. We thus impose constraints on the form of $\widehat{\corr}(|\kv|)$ to fulfill the following requirements: (i) peak in the correlation function at $|\kv|=1$ (as above); (ii) value of the peak set to 0 such to have the same phenomenological temperature/quenching depth ($B$, see also Sect.~\ref{sec:SHPFC}) as in the minimal models; (iii) same elastic constants as the triangular 1-mode (see Table \ref{tab:ge_lengths}); (iv) isotropic GE as obtained in Sect.~\ref{sec:elasge}.
As these requirements are independent, and enforcing isotropic GE imposes two additional constraints, the coefficients in Eq.~\eqref{eq:exp_corrfun} must satisfy a total of 5 conditions. Therefore, a minimum of $N=6$ is needed to tune the characteristic lengths independently, yielding
\begin{equation}
    c_2 = 2-3c_{12}, \quad
    c_4 = 1 - 9c_{12}, \quad
    c_6 = -8c_{12}, \quad
    c_8 = 0, \quad
    c_{10} = 3c_{12},
\end{equation}
which results in the following length scales, now featuring a free parameter ($c_{12}$)
\begin{equation}
    \ell_1^2 = \frac{1}{3} - 6c_{12}, \qquad
    \ell_2^2 = 1 - 6c_{12}.
\end{equation}
Note that the value and the ratio between length scales now depend on $c_{12}$. Finally, to ensure the suppression of higher frequencies and real positive characteristic lengths, $c_{12} \geq 0$ and $c_{12} \leq 1/18$ must hold, respectively, thus implying $\ell_1/\ell_2 \in [0, 1/\sqrt{3}]$. In this setting, we then obtained a SH energy functional with an extended parameterization, allowing for tuning GE constants. Following similar procedures, we envisage that extended parametrization can be devised for models based on a SH free energy encoding other lattice symmetries or coupling with other terms by considering appropriate expansions of $C_2$.

\section{Effective GE length scales at dislocations}
\label{sec:examples}

To fully characterize elasticity in the SH and PFC models, linear, nonlinear, and strain-gradient contributions must be considered. Consequently, the assumptions considered in Sect.~\ref{sec:elas-sh} to allow for the derivation of fully analytic expressions need to be relaxed.
Through numerical simulations, we show in this section that the elasticity encoded in the considered models matches well with known predictions from strain-gradient elasticity using effective characteristic lengths. These are found to be larger than the analytical predictions reported above 
and thus are in better agreement with results from the literature. Moreover, it allows us to characterize additional effects and unveil their dependence on model parameters.

\subsection{Numerical simulation of a stationary defect configuration}
\label{sec:num5}

A key evidence concerning GE at the microscale is the regularization of the elastic fields at dislocations \cite{Lazar2005}, which is also of central interest for crystalline systems. For simplicity, we consider the APFC model from Sect.~\ref{sec:apfc}, thus encoding a natural coarse-graining length in the equation (the lattice parameter \cite{salvalaglio2022coarse}). Accordingly, the elastic field can be directly computed from $\eta_j$ rather than upon numerical coarse-graining (see also Eq.~\ref{eq:sigmapsi1}).

An edge dislocation can be simulated by setting the complex amplitudes' phases as $\theta_{n}=-\kv_n\cdot \mathbf{u}^{\rm dislo}$ \cite{salvalaglio2022coarse} and letting the system evolve to equilibrium, with
\begin{equation}
    \begin{split}
        u_x^{\rm dislo}&= \frac{b}{2\pi} \bigg[ \arctan{\left(\frac{y}{x}\right)} +\frac{xy}{2(1-\nu)(x^2+y^2)} \bigg],\\
        u_y^{\rm dislo}&= -\frac{b}{2\pi} \bigg[ \frac{(1-2\nu)}{4(1-\nu)}\log{\left( x^2+y^2 \right)}+\frac{x^2-y^2}{4 (1-\nu) (x^2+y^2)} \bigg],\\
    \end{split}
    \label{eq:udislo}
\end{equation}
the dislocation displacement field, $\mathbf{b}=b_x\hat{\mathbf{x}}$ the Burgers vector, and $\nu$ the Poisson's ratio \cite{anderson2017}. In general, defect arrangements evolve dynamically \cite{SkogvollJMPS2022}. To study elastic fields at defects, we consider a static configuration. It consists of a checkered pattern, nominally on a square grid, of dislocations having $\pm \mathbf{b}$ as Burgers vectors and distance $L/2$. Notice, however, that dislocations cannot always be placed exactly on a square grid, e.g. in the presence of a triangular lattice, and the relaxation of the initial condition may lead to a small shift in their positions. However, this configuration allows simulating a static, periodic system by considering an $L \times L$ box. We remark that the displacement field in Eqs.~\eqref{eq:udislo} is exact for an isotropic medium within linear elasticity \cite{anderson2017}. Here, it is exploited to introduce dislocation with the desired Burgers vector. By letting the system relax to equilibrium, the actual solution including the effects outlined in Eq.~\eqref{eq:geta}, is obtained. Analogous results, but with a slower convergence to the equilibrium solution, can be obtained by setting proper singularities in $\theta_n$ without initial assumption on elasticity as done, e.g., in Ref.~\cite{SkaugenPRB2018}. Numerical results are obtained by exploiting a simple pseudo-spectral Fourier method as summarized in \ref{app:num}. 

\subsection{Elastic field at a dislocation}
\label{sec:elas5}

\begin{figure}
    \centering
    \includegraphics[width=1.0\textwidth]{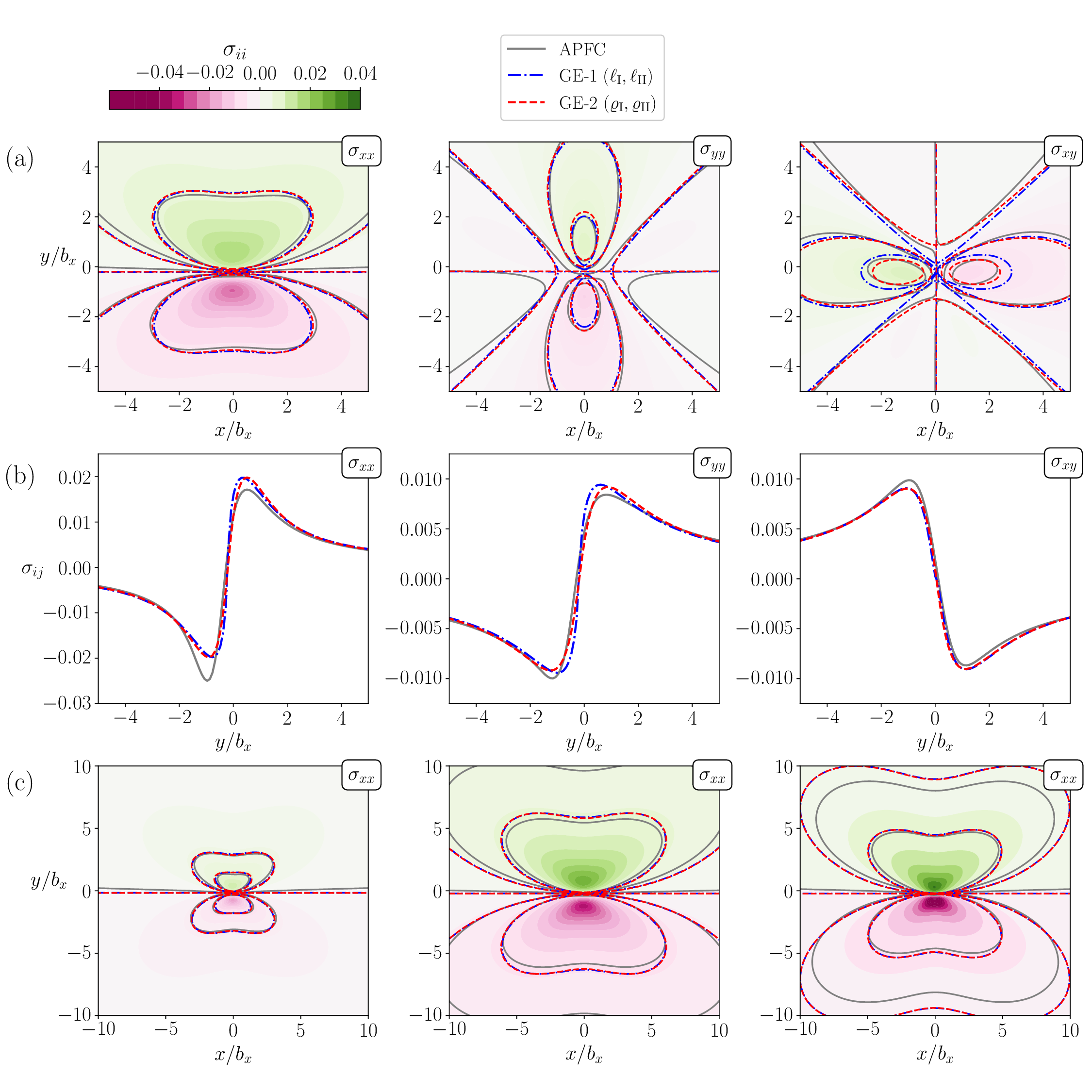}
    \caption{Stress field from APFC modeling of an edge dislocation in a triangular lattice (\textit{tri-1}) compared to the predictions of GE-1 and GE-2 for an isotropic medium. Characteristic lengths, $(\ell_{\rm I},\ell_{\rm II})$ for GE-1 and $(\varrho_{\rm I},\varrho_{\rm II})$ for GE-2, are obtained by fitting Eqs.~\eqref{eq:ge-11}--\eqref{eq:ge-13} and \eqref{eq:ge-21}--\eqref{eq:ge-23}, respectively, to the APFC numerical result (see further details and values obtained by fitting in \ref{app:fits}). APFC results are shown with solid grey contour lines and the underlying colormap to note negative and positive stress regions. The GE-1 solution with fitted length scales is illustrated via dot-dashed blue lines, and the GE-2 solution via dashed red lines. Contour lines are set to the same values in all the 2D plots of the stress to $[-7,-3.5,0.0,3.5,7]\times 10^{-3}$. The same applies to the range of color maps. (a) 2D contour plots of the stress field components $\sigma_{xx}$,  $\sigma_{yy}$, and  $\sigma_{xy}$ (from left to right) for $A=1$, $B=0.02$. (b) Detailed comparison as 1D curves of the values in panel (a) along $y$ at $x=0$ for $\sigma_{xx}$ and $\sigma_{yy}$, and along $x$ at $y=0$ for $\sigma_{xy}$ (from left to right). (c) $\sigma_{xx}$ illustrated as in panel (a) for three different parameter choices. From left to right parameters $(A,B)$ are set to: $(0.5,0.02)$, $(2.0,0.02)$ and $(1,-0.026)$. In all plots $b_x=a_0$, $\bar \psi=0$,  $C=-1/2$, and $D=1/3$.}
    \label{fig:dislo1}
\end{figure}

Figure \ref{fig:dislo1} illustrates the stress field of an edge dislocation in a triangular crystal obtained at equilibrium by APFC simulations compared to the prediction of GE theories with fitted characteristic lengths. We consider solutions for an edge dislocation in an isotropic medium obtained within GE-1 and GE-2 as reported in Refs.~\cite{LAZAR20061787} and \cite{Lazar2022}, respectively. Corresponding equations are reported in \ref{app:GEs}. Details of the fitting procedure are given in \ref{app:fits}. Fitted values of the characteristic lengths $\ell_i$ and $\varrho_i$ are indicated with the subscripts ``I" and ``II" in the following.

In particular, Fig.~\ref{fig:dislo1}a shows stress field components in 2D for a selected set of parameters. Color maps and representative grey contour lines illustrate the value obtained by numerical APFC simulations. The stress field along lines crossing the dislocation core parallel to the $y$-axis for $\sigma_{xx}$ and $\sigma_{yy}$ and the $x$-axis for $\sigma_{xy}$ are shown in Fig.~\ref{fig:dislo1}b. $\sigma_{ij}$ from APFC simulations are regularized at the core, and both GE-1 and GE-2 solutions match very well the results of numerical simulations for some values of the respective characteristic lengths ($\ell_i$ and $\varrho_i$). As observed in other works \cite{salvalaglio2022coarse}, the elastic field obtained by APFC is asymmetric, a feature that is absent in the considered GE theories as it results from nonlinearities in the free-energy functional \cite{Huter2016}. Accordingly, this effect is more pronounced for larger stress values, namely for $\sigma_{xx}$ where the stress field reaches maximum and minimum values two times larger than other components.
Characteristic lengths have been determined by fitting the numerically computed values for the stress components via the analytical solution for both GE-1 and GE-2 (reported in \ref{app:GEs}). Dislocation positions (subject to the small shift mentioned above) are retained as fitting parameters too. All the values determined via fitting for the plots in Fig.~\ref{fig:dislo1}a and Fig.~\ref{fig:dislo1}b are reported in \autoref{tab:fittable}.

While our focus here is on the prototypical case of an edge dislocation in an isotropic lattice, which allows for extended comparisons with analytical expressions and, in principle, connections to other theories, it is important to note that other dislocations (like of screw or mixed type) and lattice symmetries are also naturally described in PFC models \cite{Berry2014,Yamanaka2017,Salvalaglio2018,Skogvoll_2022,SkogvollJMPS2022}. An extended study covering more cases is out of the scope of the present work. The analysis of targeted cases is left to future applications.

\subsection{Discussion of the results}
\label{sec:discussion5}

From a quantitative point of view (see \autoref{tab:fittable}), characteristic GE-1 lengths obtained from fitting $\sigma_{xx}$ and $\sigma_{yy}$ are very similar, $\ell_{\rm I}\sim\ell_{\rm II}\sim 0.55a_0$. The values for $\sigma_{xy}$ deviate from this behavior ($\ell_{\rm I} \sim 0.23 a_0$ and $\ell_{\rm II} \sim 0.48 a_0$). Nevertheless, by using the value obtained for the other components, $\sigma_{xy}$ shows a very similar profile with respect to the maximum and minimum values as well as the decay far from the core. A small difference is observed only very close to the core, where the two significantly different lengths allow for smaller gradients and a better matching of the simulation results. When enforcing a single characteristic length ($\ell_1=\ell_2$)---which would be consistent with the conclusion of lattice theory with central force interaction between atoms \cite{Born1955,Lazar2022_iso}---the fitted values are very similar for the three components of the stress field analyzed (see the third line in \autoref{tab:fittable}). We may conclude that GE-1 can match the simulations well but with characteristic lengths that are not fully consistent among different stress-field components. Conversely, the values obtained via fitting the characteristic GE-2 lengths are very close to each other in all three stress components, assessing that the underlying model accurately describes simulation results.
By relaxing the assumption of constant amplitude moduli underlying the derivation in Sect~\ref{sec:elasge}, we indeed find that additional contributions appear, including terms entering GE-2. Further details to substantiate this argument are reported in \ref{app:ge2}.

\begin{figure}
    \centering
    \includegraphics[width=1.0\textwidth]{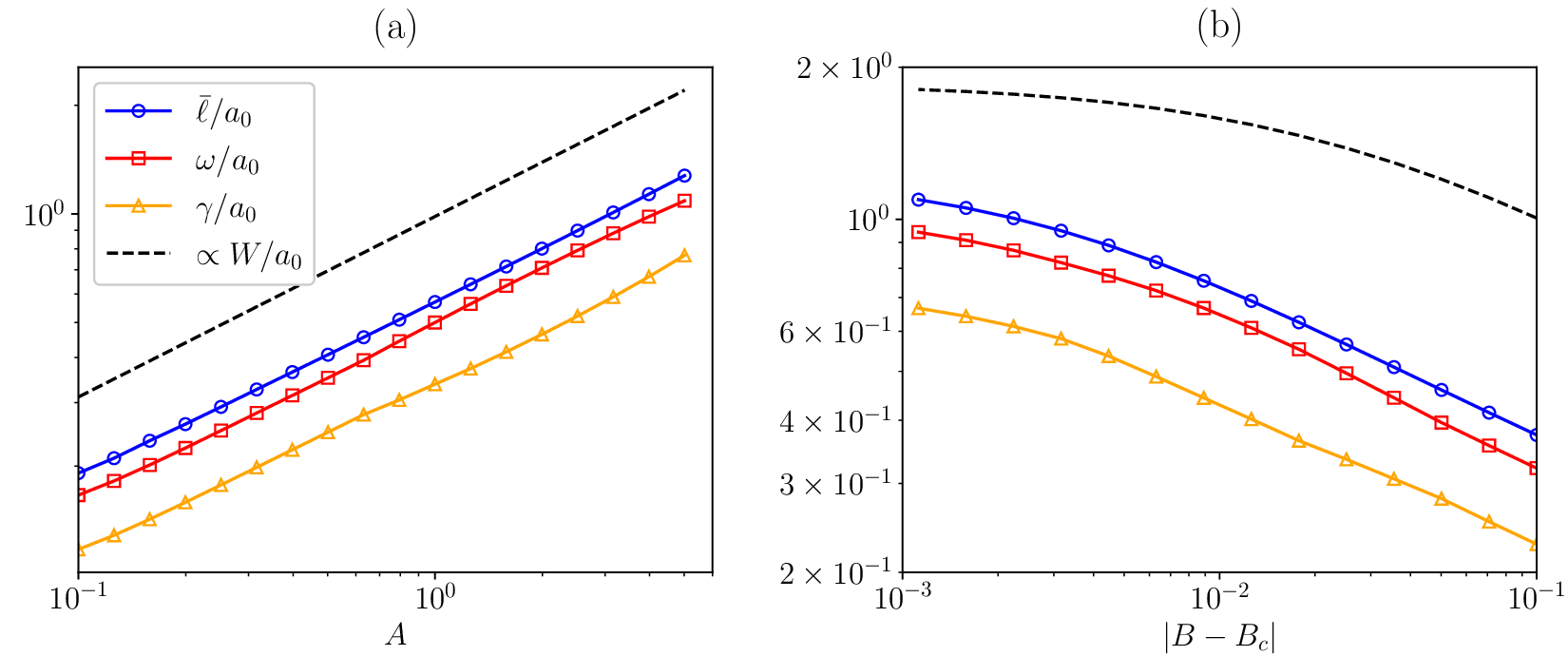}
    \caption{Dependence of the characteristic lengths on free energy parameters. Variation of characteristic lengths with (a) $A$ for $B=0.02$, and  (b) $B$ for $A=1$. In all simulations, $C=-1/2$ and $D=1/3$. The panels illustrate in particular log-log plots of $\bar{\ell}=(\ell_{\rm I}+\ell_{\rm II})/2$ (blue circles), $\omega$ (red squares) and $\gamma$ (orange triangles). Solid lines are guidelines for the eyes. Values reported here are obtained by fitting $\sigma_{yy}$. Dependence on $B$ is illustrated via the quenching depth $|B-B_c|$. The correlation length $W$ from Eq.~\eqref{eq:corrlength} is reported for comparison (dashed line).}
    \label{fig:disloAB}
\end{figure}

The agreement between the simulated and analytical stress fields holds for a broad parameter range. This is first shown in Fig.~\ref{fig:dislo1}c, via 2D stress maps and contour lines of $\sigma_{xx}$ for three simulations featuring $A$ or $B$ values different from those in Fig.~\ref{fig:dislo1}a. The change in the $\sigma_{xx}$ is reflected in a variation of the characteristic lengths determined by fitting. This aspect is further quantitatively illustrated in Fig.~\ref{fig:disloAB}, allowing for a comprehensive analysis of the considered GE theories. For this analysis we consider the characteristic lengths fitted from $\sigma_{yy}$. This component is chosen as it features an almost symmetric profile owing to lower maximum values for the stress field, and thus a limited contribution of nonlinearities, than $\sigma_{xx}$ (see also Fig.~\ref{fig:dislo1}). For GE-1, we show the average of the two characteristic lengths $\bar{\ell}=(\ell_{\rm I}+\ell_{\rm II})/2$, as their relative difference is below $5\%$ over the whole range of parameters considered (see Table \ref{tab:ge_lengths}). For GE-2 we report the two characteristic lengths $\omega=\sqrt{\varrho_{\rm  I}^2+\varrho_{\rm II}^2}$ and $\gamma=\sqrt[4]{\varrho_{\rm  I}^2\varrho_{\rm II}^2}$, allowing for direct comparisons with GE-1 (see also Sect.~\ref{sec:GE2}). We vary in particular parameters $A$ (Fig.~\ref{fig:disloAB}a) and $B$ (Fig.~\ref{fig:disloAB}b). The variation with $B$ is illustrated in terms of the quenching depth, namely $|B-B_c|$ with $B_c=8C^2/(135D)$ the value of $B$ at which ordered (periodic $\psi$ with triangular symmetry) and disordered (constant $\psi$) phases have the same energy \cite{salvalaglio2022coarse}. We note that the parameter range includes most variations typically considered for PFC models.

We generally find that effective GE characteristic lengths are larger than the one computed in Sect.~\ref{sec:elas-sh} and approach common values reported in the literature in terms of ratio with the lattice parameter \cite{shodja2013ab,Lazar2023}. Characteristic lengths of GE-1 and GE-2 vary similarly. Interestingly, $\omega \sim 0.9\bar{\ell}$. We remark that $\omega$ is the characteristic length to the lowest order of GE-2, which reduces to GE-1 in the limit $\gamma=0$. The computed value of $\gamma$ being greater than zero is thus a further demonstration of the presence of higher-order corrections from terms specific to GE-2, as supported by the argument in \ref{app:ge2}.

Importantly, the characteristic lengths are found to be proportional to the correlation length $W$, which corresponds to the width of the interface between an ordered and disordered phase, both in equilibrium and nonequilibrium settings \cite{Galenko2015}. Dashed black lines in Fig.~\ref{fig:disloAB} show the analytic expression of $W$ reported in Eq.~\eqref{eq:corrlength}, which is derived in an approximated setting similar to the ones considered in Sect.~\ref{sec:elas-sh}. Ultimately, we find a proportionality constant $\bar{\ell}/W \sim {\omega}/W \in(0.4,0.6)$. In models like SH or PFC, the extension of the defect core scales similarly with the correlation length \cite{skogvollnpj2023}. Therefore, the effective characteristic length---which we recall encodes microscopic effects into (continuum) elasticity---is found to scale \textit{de facto} with the core size, consistently with continuum descriptions of the elastic field of dislocations in GE theories \cite{Lazar2005,LAZAR20061787,Lazar2022}. By recalling that $B$ is a phenomenological temperature parameter, the effective GE characteristic lengths at dislocations are then connected to the temperature.


\section{Conclusions}
\label{sec:conclusions}

We discussed how gradient elasticity (GE) is encoded in Swift-Hohenberg (SH) and phase-field crystal (PFC) models. The leading order for small deformations consists of a first strain-gradient elasticity formulation, anisotropic for stripe phases and isotropic for crystalline phases. We derived analytical or semi-analytical formulations for GE characteristic lengths for one and multimode expansions of the periodic order parameter for different crystalline arrangements in an approximated setting. 

These characteristic lengths are in the order of fractions of the lattice parameter ($\sim$1\r{A}) as observed for several materials. From a quantitative point of view, they are consistent with results reported in the literature. However, they underestimate the ones commonly found for real materials with the same lattice symmetries \cite{shodja2013ab,Lazar2023}. When considering the minimal formulation of the SH energy functional, they can only be slightly varied, conforming to the known restrictions on tuning elastic constants in these approaches compared to fully atomistic methods. However, we outlined the possibility of extending the parametrization via the design of the nonlocal terms in the free energy. Although these results outline the leading-order GE effects, they are obtained via simplifying assumptions, namely neglecting nonlinear terms and assuming constant amplitude moduli everywhere.

Numerical simulations that allow the inspection of GE effects without approximations show that larger effective GE characteristic lengths emerge. In addition, they can be significantly varied with the model parameters. We focused on the elastic field at dislocations, which is of central interest for both GE and SH/PFC theories. Interestingly, the analytical solutions of stress fields at edge dislocations in both first and second strain-gradient elasticity closely resemble the results of APFC simulations. By analyzing the values of the emerging effective characteristic lengths, we found GE-2 to be a better model for describing the simulated stress field, thus assessing the underlying GE theory. Overall, we showed that these GE theories, particularly GE-2, naturally emerge from the minimal framework of the SH energy functional, which can be considered the simplest free-energy form minimized by a periodic, smooth order parameter \cite{Elder2002}.

The effective GE characteristic lengths at dislocations vary with the model parameters similarly to the phase correlation length. This closely resembles variations observed for the size of defect cores in smooth theories for ordered systems \cite{skogvollnpj2023}. Importantly, the variation in the quenching depth can be interpreted as a dependence on temperature, which may constitute the input for other theories and establishes a direct link between GE and the PFC/SH framework for order-disorder phase transition.

\section*{Acknowledgements}
We acknowledge interesting discussions with Ken R. Elder, Markus Lazar, and Axel Voigt. We also gratefully acknowledge support from the German Research Foundation under Grant No. SA4032/2 -- Emmy Noether Programme -- (MS) and SA4032/3 -- FOR3013 -- (LBM), and from the Center for Information Services and High-Performance Computing [Zentrum für Informationsdienste und Hochleistungsrechnen (ZIH)] at TU Dresden for computing time.

\section*{Data Availability Statement}
The data that support the findings of this study are openly available at the following URL/DOI: \href{https://www.doi.org/10.5281/zenodo.10351669}{\texttt{10.5281/zenodo.10351669}} 

\appendix

\section{Numerical method}
\label{app:num}
Numerical simulations are computed by exploiting a Fourier pseudo-spectral method. In brief, we solve the equation(s) \eqref{eq:dynAPFC} rewritten as
\begin{equation}
\partial_t{\eta_n} = \mathfrak{L} \eta_n + \mathfrak{N}(\{\eta_n\}),
\end{equation}
with $\eta_n$ the amplitudes to solve for, $\mathfrak{L}$ and $\mathfrak{N}$  the linear operator and polynomial terms in $\eta_n$ of Eq.~\eqref{eq:dynAPFC}, respectively. We recall that $\mathfrak{N}$ includes nonlinear terms in all the amplitudes $\eta_n$. In the Fourier pseudo-spectral method, we then solve for
\begin{equation}
\partial_t{[\widehat{\eta}_n]_k} = \widehat{\mathfrak{L}}_k [\widehat{\eta}_n]_k + \widehat{\mathfrak{N}}_k,
\label{eq:ode}
\end{equation}
with $[\widehat{\eta}_n]_k$ the coefficient of the Fourier transform of ${\eta}_n$, $\widehat{\mathfrak{N}}_k$ the Fourier transform of $\mathfrak{N}(\{\eta_n\})$ and $\widehat{\mathfrak{L}_k}$ the Fourier transform of $\mathfrak{L}$ resulting in an algebraic expression of the wave vector (for instance, in 1D, for $\mathfrak{L}\eta =\frac{\partial^2 \eta}{\partial x^2}$ one gets $\widehat{\mathfrak{L}}_k=-k^2$ with $k$ the coordinate in the Fourier space). The solution at $t+\Delta t$, with $\Delta t$ the timestep, is then obtained via an inverse Fourier transform of $[\widehat{\eta}_n]_k(t+\Delta t)$ computed by the following approximation
\begin{equation}
[\widehat{\eta}_n]_k(t+\Delta t) 
\approx\ {\rm e}^{\mathfrak{L}_k\Delta t}[\widehat{\eta}_n]_k(t)
+\frac{{\rm e}^{\mathfrak{L}_k\Delta t}-1}{\mathfrak{L}_k} \widehat{\mathfrak{N}}_k(t).
\label{eq:spec}
\end{equation}
This method enforces periodic boundary conditions.
We implemented it in \texttt{python} (code openly available, see \cite{apfc-code}). An established Fast-Fourier Transform algorithm (FFTW) is exploited \cite{Frigo2005}. We use a discretization of 4 mesh points per atomic site with a timestep $\Delta t = 1$.

From $\{\eta_n\}$ computed with the method outlined above, we compute the stress field components. This is achieved via the equation \cite{SalvalaglioJMPS2020}
\begin{equation}
\begin{split}
    \sigma_{ij}=\sum_{n=1}^N&
    \bigg\{ \big[(\partial_i + {\rm i} (\kv_n)_i)(\nabla^2 + 2{\rm i}\mathbf{k}_n\cdot \nabla)\eta_n \big] \big[(\partial_j - {\rm i} (\kv_n)_j)\eta_n^*\big] \\
    &- \big[(\nabla^2 + 2{\rm i}\mathbf{k}_n\cdot \nabla) \eta_n\big] \big[(\partial_i - {\rm i} (\kv_n)_i)(\partial_j - {\rm i} (\kv_n)_j)\eta_n^*+{\rm c.c.}\big] 
    \bigg\}.
    \label{eq:sigmapsi1}
\end{split}
\end{equation}

\section{Stress field of an edge dislocation in strain-gradient elasticity}
\label{app:GEs}

In Toupin-Mindlin first strain-gradient elasticity (GE-1, Sect.~\ref{sec:GE1}), the stress field in the $xy$-plane for an edge dislocation located at the origin with Burgers vector aligned with the $x$-axis, $\mathbf{b}=b_x\hat{\mathbf{x}}$,  reads \cite{Lazar2022}:
\begin{equation}\label{eq:ge-11}
\begin{split}
    \sigma_{xx} = -\sigma_0 \frac{y}{r^2}&\left\{\frac{3x^2+y^2}{r^2} - \frac{2\nu r}{\ell_1}K_1(r/\ell_1) \right. \\
    &+(1-2\nu)\left[\frac{3x^2-y^2}{r^2}\left(\frac{4\ell_1^2}{r^2}-2K_2(r/\ell_1)\right) - \frac{2x^2}{\ell_1 r}K_1(r/\ell_1)\right] \\
    &-2(1-\nu)\left[\frac{3x^2-y^2}{r^2}\left(\frac{4\ell_2^2}{r^2}-2K_2(r/\ell_2)\right) - \frac{x^2-y^2}{\ell_2 r}K_1(r/\ell_2)\right], \\
\end{split}
\end{equation}
\begin{equation}\label{eq:ge-12}
\begin{split}
    \sigma_{yy} = \sigma_0 \frac{y}{r^2}&\left\{\frac{x^2-y^2}{r^2} + \frac{2\nu r}{\ell_1}K_1(r/\ell_1) \right. \\
    &+(1-2\nu)\left[\frac{3x^2-y^2}{r^2}\left(\frac{4\ell_1^2}{r^2}-2K_2(r/\ell_1)\right) + \frac{2y^2}{\ell_1 r}K_1(r/\ell_1)\right] \\
    &-2(1-\nu)\left[\frac{3x^2-y^2}{r^2}\left(\frac{4\ell_2^2}{r^2}-2K_2(r/\ell_2)\right) - \frac{x^2-y^2}{\ell_2 r}K_1(r/\ell_2)\right], \\
\end{split}
\end{equation}
\begin{equation}\label{eq:ge-13}
\begin{split}
    \sigma_{xy} = \sigma_0 \frac{x}{r^2}&\left\{\frac{x^2-y^2}{r^2} \right. \\
    &+(1-2\nu)\left[\frac{x^2-3y^2}{r^2}\left(\frac{4\ell_1^2}{r^2}-2K_2(r/\ell_1)\right) + \frac{2y^2}{\ell_1 r}K_1(r/\ell_1)\right] \\
    &-2(1-\nu)\left[\frac{x^2-3y^2}{r^2}\left(\frac{4\ell_2^2}{r^2}-2K_2(r/\ell_2)\right) + \frac{2y^2}{\ell_2 r}K_1(r/\ell_2)\right],
    \end{split}
\end{equation}
with $\sigma_0=(\mu b_x)/(2 \pi(1-\nu))$, $r=\sqrt{x^2+y^2}$, and $K_n(x)$ are the modified Bessel functions of the second kind ($n=1,2$ here). 
The classical (singular) stress field obtained in linear elasticity is recovered for $\ell_1=\ell_2=\ell \rightarrow 0$ \cite{anderson2017}.

In the special second strain-gradient elasticity theory (GE-2) outlined in Sect.~\ref{sec:GE2}, the stress field components for the same dislocation read \cite{LAZAR20061787} 
\begin{equation} \label{eq:ge-21}
\begin{split}
\sigma_{x x}=  -\sigma_0 \frac{y}{r^4}\bigg\{ & \left(y^2+3 x^2\right)+\frac{4\left(\varrho_1^2+\varrho_2^2\right)}{r^2}\left(y^2-3 x^2\right) -\frac{2 r y^2}{\varrho_1^2-\varrho_2^2}\left[\varrho_1  K_1\left(\frac{r}{\varrho_1}\right)-\varrho_2  K_1\left(\frac{r}{\varrho_2}\right)\right] \\ & -\frac{2\left(y^2-3 x^2\right)}{\varrho_1^2-\varrho_2^2}\left[\varrho_1^2 K_2\left(\frac{r}{\varrho_1}\right)-\varrho_2^2 K_2\left(\frac{r}{\varrho_2}\right)\right]\bigg\},
\end{split}
\end{equation}
\begin{equation} \label{eq:ge-22}
\begin{split}
\sigma_{y y}=  -\sigma_0\frac{y}{r^4}\bigg\{& \left(y^2-x^2\right)-\frac{4\left(\varrho_1^2+\varrho_2^2\right)}{r^2}\left(y^2-3 x^2\right)-\frac{2 r x^2}{\varrho_1^2-\varrho_2^2}\left[\varrho_1  K_1\left(\frac{r}{\varrho_1}\right)-\varrho_2  K_1\left(\frac{r}{\varrho_2}\right)\right] \\
& +\frac{2\left(y^2-3 x^2\right)}{\varrho_1^2-\varrho_2^2}\left[\varrho_1^2 K_2\left(\frac{r}{\varrho_1}\right)-\varrho_2^2 K_2\left(\frac{r}{\varrho_2}\right)\right]\bigg\},
\end{split}
\end{equation}
\begin{equation} \label{eq:ge-23}
\begin{split}
\sigma_{x y}= \sigma_0 \frac{x}{r^4}\bigg\{&\left(x^2-y^2\right)-\frac{4\left(\varrho_1^2+\varrho_2^2\right)}{r^2}\left(x^2-3 y^2\right)-\frac{2 r y^2}{\varrho_1^2-\varrho_2^2}\left[\varrho_1  K_1\left(\frac{r}{\varrho_1}\right)-\varrho_2  K_1\left(\frac{r}{\varrho_2}\right)\right] \\
& +\frac{2\left(x^2-3 y^2\right)}{\varrho_1^2-\varrho_2^2}\left[\varrho_1^2 K_2\left(\frac{r}{\varrho_1}\right)-\varrho_2^2 K_2\left(\frac{r}{\varrho_2}\right)\right]\bigg\}.
\end{split}
\end{equation}

\begin{figure}[h!]
    \centering
    \includegraphics[width=1.0\textwidth]{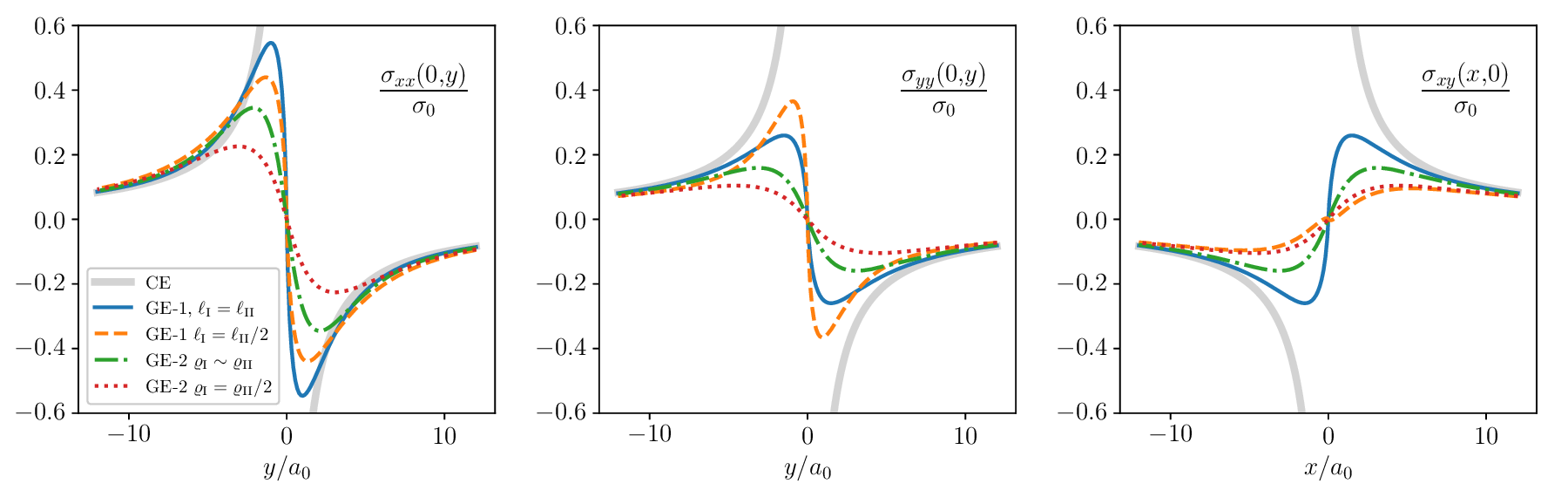}
    \caption{Stress field components $\sigma_{xx}$, $\sigma_{yy}$, and $\sigma_{xy}$ (left to right) from GE-1 and GE-2, namely Eqs.~\eqref{eq:ge-11}--\eqref{eq:ge-13} \cite{Lazar2022} and \eqref{eq:ge-21}--\eqref{eq:ge-23} \cite{LAZAR20061787}, respectively. Two ratios for characteristic lengths are showcased, where $\ell_1=\varrho_1=1$. The linear continuum elasticity solution (CE) \cite{anderson2017} is obtained with Eqs.~\eqref{eq:ge-11}--\eqref{eq:ge-13} in the limit $\ell_1=\ell_2 \rightarrow 0$.}
    \label{fig:disloGE}
\end{figure}

An illustration of the stress field components for GE-1, GE-2, and the continuum elasticity limit CE, is reported in Fig.~\ref{fig:disloGE}.

\section{Details of fitting procedures for characteristic lengths}
\label{app:fits}

The equations reported in \ref{app:GEs}, upon centering the frame of reference at the dislocation core in $(x_0, y_0)$, illustrate functions $\sigma_{ij}(x,y,x_0,y_0,E,\nu,\xi_1,\xi_2)$ with $\xi_i=\ell_i$ for GE-1 and $\xi_i=\varrho_i$ for GE-2. They are fitted (exploiting \texttt{scipy.optimize} \cite{2020SciPy-NMeth}) on $\sigma_{ij}(x,y)$ in a squared region $15a_0 \times 15a_0$ centered on a defect to determine $\xi_i$. Exploiting the elastic constants obtained in \autoref{tab:ge_lengths}, and in agreement with \cite{ElderPRE2010,HeinonenPRE2014}, we have that for the one-mode approximation of the triangular lattice the Lam\'e constants read $\mu=\lambda=3 A \varphi_1^2$, which results in the Young modulus $E=5 A \varphi_1^2/2$ and Poisson ratio $\nu=1/4$ (under the plane strain assumption). $\varphi$ can be computed analytically for one-mode approximations \cite{salvalaglio2022coarse}. The value for the considered triangular lattice is $\varphi=(-C+\sqrt{C^2-15BD})/(15D)$. The dislocation in the APFC simulations moves slightly upon reaching equilibrium. In general, $x_0$ and $y_0$ must be considered as fitting parameters as well. The fit results for the fields illustrated in Fig.~\ref{fig:dislo1} are reported in \autoref{tab:fittable}. The fitted values of the characteristic lengths $\ell_i$ and $\varrho_i$ are indicated with the subscripts ``I" and ``II".

\begin{table}[ht]
\renewcommand{\arraystretch}{1.5}
\centering
\begin{tabular}{@{}c|ccc|c@{}}
\toprule
\  \ & 
\ $\sigma_{xx}$   \          & 
\ $\sigma_{yy}$ \    &
\ $\sigma_{xy}$ \    &
\ Average \ 
\\ \midrule
$\ell_{\rm I}$ \quad  & 
{\small $0.5413$} & 
{\small $0.5600$} & 
{\small $0.2275$} & 
{\small $0.4429$}
\\
$\ell_{\rm II}$ \quad  & 
{\small $0.5800$} & 
{\small $0.5789$} & 
{\small $0.4303$} & 
{\small $0.5290$}
\\
$\ell_{\rm I}=\ell_{\rm II}$ \quad  & 
{\small $0.5630$} & 
{\small $0.5597$} & 
{\small $0.5692$}  
& {\small $0.5640$}
\\
$\varrho_{\rm I}$ \quad  & 
{\small $0.4472$} & 
{\small $0.4227$} & 
{\small $0.4275$} 
& {\small $0.4325$}
\\
$\varrho_{\rm II}$ \quad  & 
{\small $0.2415$} & 
{\small $0.2673$} & 
{\small $0.2746$}  
& {\small $0.2612$}
\\
$x_0^{\text{GE-1}}$ \quad  & 
{\small $-0.0216$} & 
{\small $ 0.0072$} & 
{\small $ 0.0450$}  
& {\small $0.0101$}
\\
$x_0^{\text{GE-2}}$  \quad  & 
{\small $-0.0204$} & 
{\small $0.0065$} & 
{\small $0.0452$}  
& {\small $0.0104$}
\\
$y_0^{\text{GE-1}}$  \quad  & 
{\small $-0.2079$} & 
{\small $-0.1951$} & 
{\small $-0.2088$}  
& {\small $-0.2039$}
\\
$y_0^{\text{GE-2}}$ \quad  & 
{\small $-0.2000$} & 
{\small $-0.1839$} & 
{\small $-0.21103$}  
& {\small $-0.1983$}
\\
\bottomrule
\end{tabular}
\caption{Results of the fits illustrated in Fig.~\ref{fig:dislo1}(a). Values $(\ell_{\rm I},\ell_{\rm II},x_0^{\text{GE-1}},y_0^{\text{GE-1}})$ are obtained by fitting Eqs~\eqref{eq:ge-11}--\eqref{eq:ge-13} with characteristic lengths and positions as fitting parameters. The values $\ell_{\rm I}=\ell_{\rm II}$ are obtained from fitting the same equation but enforcing equal characteristic lengths. Values $(\varrho_{\rm I},\varrho_{\rm II},x_0^{\text{GE-2}},y_0^{\text{GE-2}})$ are obtained by fitting Eqs~\eqref{eq:ge-21}--\eqref{eq:ge-23}, also in this case with characteristic lengths and positions as fitting parameters. All quantities are expressed in $a_0$ units (fraction of the lattice parameter). For all these values, the standard deviation does not exceed $2 \times 10^{-4}$.}
\label{tab:fittable}
\end{table}

The values of fitted parameters are overall consistent across different stress components, with just GE-1 characteristic lengths obtained by fitting $\sigma_{xy}$ deviating from other components, as discussed in the main text. Note that $x_0$ differs from $0$ by less than $5\%$ of the lattice parameter. The shift along the $x$ direction could then be considered negligible. Instead, we obtain a shift along $y$ of $\sim 20\%$ of the lattice parameter. Note that the dislocation core location $(x_0,y_0)$ is quantitatively consistent across the fitted values on both GE-1 and GE-2 stress field components, where indeed $x_0$ and $y_0$ have the same meaning.
In the plots reported in the main text, the stress field from GE solutions for dislocations are centered in the $(x_0,y_0)$ values obtained from fitting the corresponding stress component. We remark that in GE-2, the two length scales $\varrho_{1,2}$ are fully equivalent as it follows from Eq.~\eqref{eq:mechge-2}, so their values could be exchanged. Also, in some parameter ranges, e.g., for small $A$ values, the best fit is given by $\varrho_{1}\sim\varrho_{2}$ while in other cases a significant difference, as in Table \ref{tab:fittable} above, is obtained. The physically meaningful lengths entering the theory are, however, $\omega(\varrho_{1},\varrho_{2})$ and $\gamma(\varrho_{1},\varrho_{2})$ as defined in Sect.~\ref{sec:GE2} and these are the quantities illustrated in Fig.~\ref{fig:disloAB}.
This figure also reports the correlation length $W$ for triangular lattice when varying $A$ and $B$. For this lattice symmetry in the one-mode approximation, an analytic solution connecting $W$ to the model parameter exists \cite{Galenko2015}
\begin{equation}\label{eq:corrlength}
W(A,B)=\sqrt{\frac{A}{|B|}}\frac{4\sqrt{2}}{\zeta+\sqrt{\zeta^2-4\iota}},
\end{equation}
with $\iota=\mp 1$ for $B \lessgtr 0$ and $\zeta=2C/\sqrt{15|B|D}$.

\section{On higher order GE terms from nonconstant amplitudes}
\label{app:ge2}

In Sect.~\ref{sec:elas-sh}, we show that under the assumption of constant magnitude of the amplitudes ($|\nabla\varphi_n| = 0$) typically considered for small deformations, GE is encoded in the SH energy functional via first strain-gradient terms only. This assumption, however, does not strictly hold in numerical simulations \cite{Huter2016,Wang2018,Ainsworth2019}. While refraining from a complete derivation of the theory, we report here a simplified argument to show how including the amplitude gradients leads to second strain-gradient terms in the elastic energy.

In one spatial dimension, by assuming that the gradient of the amplitude is of the same order as the gradient of the displacement while still ignoring all higher-order nonlinear terms as in Sect.~\ref{sec:SHPFC}, the elastic energy density reads:
\begin{equation}
\begin{aligned}
    \frac{w}{A'} =& k^2[4k^2u'(x)^2 + u''(x)^2]\varphi(x) + 4k^2[\varphi'(x) - \varphi(x)u''(x)]\varphi'(x)
    \\
    &+[\varphi''(x)+4k^2\varphi(x)u'(x)]\varphi''(x).
\end{aligned}
\end{equation}
Minimizing this expression with respect to the amplitude $\varphi(x)$, we obtain:
\begin{equation}
\begin{aligned}
    &\varphi''(x) = \bigg\{\frac{B+3D\varphi(x)^2}{A'(1+4k^2)}
    + \frac{A'k^2}{A'(1+4k^2)}\left[4k^2u'(x)^2 - (2-u''(x))u''(x) + 2 u'''(x)\right]\bigg\}\varphi(x).
\end{aligned}
\end{equation}
Therefore, $\varphi''(x)$ and, by extension, the elastic energy density, contain a contribution from $u'''(x)$, i.e. the second derivative of the strain field, which is the higher-order term entering GE-2 but absent in GE-1. As such, GE-2 effects may be ascribed to gradients in the amplitude, while their presence is  
anyhow corroborated by numerical results and, importantly, a detailed comparison with specific analytical solutions as discussed in Sect.~\ref{sec:examples}.

\section*{References}
\providecommand{\newblock}{}

\end{document}